%% file: maps-main.tex
\documentclass[12pt]{article}
\title{Multi Anchor Point Shrinkage for the Sample Covariance Matrix (Extended Version)}
\author{Hubeyb Gurdogan, Alec Kercheval}
\date{}
\usepackage{comment}
\usepackage{amsthm}
\usepackage{amsmath}
\usepackage{amssymb}
\usepackage{fancyhdr}
\usepackage{graphicx}
\usepackage{natbib}
\usepackage{caption}
\usepackage{subcaption}
\usepackage{morefloats}
\usepackage{float}
\usepackage{perpage}
\MakeSorted{figure}
\MakeSorted{table}
\newtheorem{thm}{Theorem}[section]
\newtheorem{lem}[thm]{Lemma}
\newtheorem{prop}[thm]{Proposition}

\theoremstyle{definition}
\newtheorem{defn}{Definition}[section]

\theoremstyle{remark}

\usepackage{enumerate}

\setcitestyle{open={[},close = {]}}
\bibpunct{[}{]}{;}{a}{}{,}

\def\bR{\mathbb{R}} 

\def\bE{\mathbb{E}}

\begin{document}

\maketitle

\begin{center}
    Last revision:  September 9, 2021
\end{center}

\begin{abstract}

Portfolio managers faced with limited sample sizes must use factor models to estimate the covariance matrix of a high-dimensional returns vector. For the simplest one-factor market model, success rests on the quality of the estimated leading eigenvector ``beta". 

When only the returns themselves are observed, the practitioner has available the ``PCA" estimate equal to the leading eigenvector of the sample covariance matrix. This estimator performs poorly in various ways.  To address this problem in the high-dimension, limited sample size asymptotic regime and in the context of estimating the minimum variance portfolio, Goldberg, Papanicolau, and Shkolnik (\cite{goldberg2018}) developed a shrinkage method (the ``GPS estimator") that improves the PCA estimator of beta by shrinking it toward the target unit vector
$q = (1,\dots, 1)/\sqrt{p} \in \bR^p$.

In this paper we continue their work to develop a more general framework of shrinkage targets that allows the practitioner to make use of further information to improve the estimator. Examples include sector separation of stock betas, and recent information from prior estimates.  We prove some precise statements and illustrate the resulting improvements over the GPS estimator with some numerical experiments.

\bigskip

Acknowledgement:   The authors thank Lisa Goldberg and Alex Shkolnik for
 helpful conversations. Any remaining errors are our own.
\end{abstract}

\tableofcontents


\section{Introduction}
\label{sec:intro}
\input{maps-intro-extended}

\section{Main Theorems}
\label{sec:maintheorems}
\input{maps-maintheorems}

\section{Tracking Error}
\label{sec:trackingerror}
\input{maps-trackingerror}

\section{Simulation Experiments}
\label{sec:simulations}
\input{maps-simulations}

\section{Proofs of the Main Theorems}
\label{sec:oracletheorems}

\input{maps-mainproofs-extended}

\section{Supplemental Proofs}
\label{sec:supplement}
\input{tech-proofs}

\bibliographystyle{apalike}
\bibliography{biblio}

\end{document}

%% file: maps-intro-extended.tex

This paper is about the problem of estimating covariance matrices for large random vectors, when the data for estimation is a relatively small sample. We discuss a shrinkage approach to reducing the sampling error asymptotically in the high dimensional, bounded sample size regime, denoted HL. We note at the outset that this context differs from that of the more well-known random matrix theory of the asymptotic ``HH regime" in which the sample size grows in proportion to the dimension (e.g. \cite{karoui2008}). See \cite{hall2005} for earlier discussion of the HL regime, and \cite{fan2008} for a discussion of the estimation problem for factor models in high dimension.

Our interest in the HL asymptotic regime comes from the problem of portfolio optimization in financial markets.  There, a portfolio manager is likely to confront a large number of assets, like stocks, in a universe of hundreds or thousands of individual issues.  However, typical return periods of days, weeks, or months, combined with the irrelevance of the distant past, mean that the useful length of data time series is usually much shorter than the dimension of the returns vectors being estimated.

In this paper we extend the successful shrinkage approach introduced in \cite{goldberg2018} (GPS) to a framework that allows the user to incorporate additional information into the shrinkage target and improve results. Our ``multi anchor point shrinkage" (MAPS) approach includes the GPS method as a special case, but improves results when some {\it a priori} order information about the betas is known.

The problem of sampling error for portfolio optimization has been widely studied ever since \cite{markowitz1952} introduced the approach of mean-variance optimization.  That paper immediately gave rise to 
the importance of estimating the covariance matrix $\Sigma$ of asset returns, as the risk, measured by variance of returns, is given by $w^{T} \Sigma w$, where $w$ is the vector of weights defining the portfolio.

 For a survey of various approaches over the years, see \cite{goldberg2018} and references therein. Reducing the number of parameters via factor models has long been standard; see for example \cite{rosenberg1974} and \cite{ross1976}.  \cite{vasicek1973} and \cite{frost1986} initiated a Bayesian approach to portfolio estimation and the efficient frontier. Vasicek used a prior cross-sectional distribution for betas to produce an empirical Bayes estimator for beta that amounts to shrinking the least-squares estimator toward the prior in an optimal way. This is one of a number of ``shrinkage" approaches in which initial sample estimates of the covariance matrix are ``shrunk" toward a prior e.g. \cite{lai2008}, \cite{bickel2008}, \cite{ledoit2003}, \cite{ledoit2004}, \cite{fan2013}.  \cite{ledoit2017} describes a nonlinear shrinkage of the covariance matrix focused on correcting the eigenvalues, set in the HH asymptotic regime.

The key insight of \cite{goldberg2018} was to identify the PCA leading eigenvector of the sample covariance matrix as the primary culprit contributing to sampling error for the minimum variance portfolio problem in the HL asymptotic regime. Their approach to {\em eigenvector} shrinkage is not explicitly Bayesian, but can be viewed in that spirit.  This is the starting point for the present work.

\subsection{Mathematical setting and background}

Next we describe the mathematical setting, motivation, and results in more detail. We restrict attention to a familiar and well-studied baseline model for financial returns: the one-factor, or ``market", model
\begin{equation} \label{eq:model}
\mathbf{r} = \beta x + \mathbf{z},
\end{equation}
where $\mathbf{r} \in \bR^p$ is a $p$-dimensional random vector of asset (excess) returns in a universe of $p$ assets, $\beta \in \bR^p$ is an unobserved non-zero vector of parameters to be estimated, $x$ is an unobserved random variable representing the common factor return, and $\mathbf{z} \in \bR^p$ is an unobserved random vector of residual returns. 

With the assumption that the components of $\mathbf{z}$ are uncorrelated with  $x$ and each other,  the returns of different assets are correlated only through $\beta$, and therefore the covariance matrix of $\mathbf{r}$ is
\begin{equation*} 
\Sigma = \sigma^2 \beta \beta^T + \Delta,
\end{equation*}
where $\sigma^2$ denotes the variance of $x$, and $\Delta$ is the diagonal covariance matrix of $\mathbf{z}$.

Under the further simplifying model assumption\footnote{The assumption of homogeneous residual variance $\delta^2$ is a mathematical convenience.  If the diagonal covariance matrix $\Delta$ of residual returns can be reasonably estimated, then the problem can be rescaled as $\Delta^{-1/2}\mathbf{r} = \Delta^{-1/2}\beta x + \Delta^{-1/2} \mathbf{z}$, which has covariance matrix $\sigma^2 \beta_{\Delta} \beta_{\Delta}^T + I$, where $\beta_{\Delta} = \Delta^{-1/2} \beta$. }
that each component of $\mathbf{z}$ has a common variance $\delta^2$ (also not observed), we obtain the covariance matrix of returns
\begin{equation} \label{eq:covariance-model}
\Sigma = \sigma^2 \beta \beta^T + \delta^2 \mathbf{I},
\end{equation}
where $\mathbf{I}$ denotes the $p \times p$ identity matrix.

This means that $\beta$, or its normalization $b = \beta/||\beta||$, is the leading eigenvector of $\Sigma$, corresponding to the largest eigenvalue $\sigma^2 ||\beta||^2 + \delta^2$.  As estimating $b$ becomes the most significant part of the estimation problem for $\Sigma$, a natural approach is to take as an estimate the first principal component (leading unit eigenvector) $h_{PCA}$ of the sample covariance of returns data generated by the model.  
This principal component analysis (PCA) estimate is our starting point.

Consider the optimization problem
\begin{eqnarray*}
\min_{w \in \bR^p} w^T \Sigma w \\
e^Tw = 1
\end{eqnarray*}
where $e=(1,1, \dots, 1)$, the vector of all ones.

The solution, the ``minimum variance portfolio", is the unique fully invested portfolio minimizing the variance of returns.  Of course the true covariance matrix $\Sigma$ is not observable and must be estimated from data.  Denote an estimate by
\begin{equation}
\hat \Sigma = \hat \sigma^2 \hat \beta \hat \beta^T + \hat \delta^2 \mathbf{I}
\end{equation}
corresponding to estimated parameters $\hat \sigma$, $\hat \beta$, and $\hat \delta$.

Let $\hat w$ denote the solution of the optimization problem
\begin{eqnarray*}
\min_{w \in \bR^p} w^T \hat \Sigma w \\
e^Tw = 1.
\end{eqnarray*}
It is interesting to compare the estimated minimum variance
\[
\hat V^2 = \hat w^T \hat \Sigma \hat w
\]
with the actual variance of $\hat w$:
\[
V^2 = \hat w^T \Sigma \hat w,
\]
and consider the variance forecast ratio $V^2/\hat V^2$ as one measure of the error made in the estimation of minimum variance, hence of the covariance matrix $\Sigma$.

The remarkable fact proved in \cite{goldberg2018} is that, asymptotically as $p$ tends to infinity, the true variance of the estimated portfolio doesn't depend on $\hat \sigma$, $\hat \delta$, or
$||\hat \beta||$, but only on the unit eigenvector
$\hat \beta/||\hat \beta||$.  Under some mild assumptions stated later, they show the following.
\begin{defn}
For a $p$-vector $\beta = (\beta(1),\dots,\beta(p))$, define the mean $\mu(\beta)$ and dispersion $d^2(\beta)$ of $\beta$ by
\begin{equation}
\mu(\beta) = \frac{1}{p} \sum_{i=1}^p \beta(i) \text{ and } d^2(\beta) = \frac{1}{p} \sum_{i=1}^p \big( \frac{\beta(i)}{\mu(\beta)} -1 \big)^2.
\end{equation}
\end{defn}

We use the notation for normalized vectors
\[
b = \frac{\beta}{||\beta||}, \text{ } q = \frac{e}{\sqrt{p}}, \text{ and } h = \frac{\hat \beta}{||\hat \beta||}.
\]
\begin{prop}[\cite{goldberg2018}] \label{prop:gpsvariance}
The true variance of the estimated portfolio $\hat w$ is given by
\[
V^2 = \hat w^T \Sigma \hat w = \sigma^2 \mu^2(\beta) (1 + d^2(\beta)) \mathcal{E}^2(h) + o_p
\]
where
$\mathcal{E}(h)$ is defined by
\[
\mathcal{E}(h) = \frac{(b,q) - (b,h)(h,q)}{1-(h,q)^2},
\]
and where the remainder $o_p$ is such that for some constants $c, C$, $c/p \leq o_p \leq C/p$ for all $p$ sufficiently large.

In addition, the variance forecast ratio 
$
V^2/\hat V^2
$
is asymptotically equal to $p \mathcal{E}^2(h)$.

\end{prop}
Goldberg, Papanicolaou and Shkolnik call the quantity $\mathcal{E}(h)$ the {\em optimization bias} associated to an estimate $h$ of the true vector $b$. They note that the optimization bias $\mathcal{E}(h_{PCA})$ is asymptotically bounded above zero almost surely, and hence  the variance forecast ratio explodes as $p \to \infty$.

With this background, the estimation problem becomes focused on finding a better estimate $h$ of $b$ from an observed time series of returns. GPS \cite{goldberg2018} introduces a shrinkage estimate for $b$ -- the GPS estimator $h_{GPS}$ -- obtained by ``shrinking" the PCA eigenvector $h_{PCA}$ along the unit sphere toward $q$, to reduce excess dispersion.
That is, $h_{GPS}$ is obtained by moving a specified distance (computed only from observed data) toward $q$  along the spherical geodesic connecting $h_{PCA}$ and $q$. ``Shrinkage" refers to the reduced geodesic distance to the ``shrinkage target" $q$.

The GPS estimator $h_{GPS}$ is a significant improvement on $h_{PCA}$. First,  $\mathcal{E}(h_{GPS})$ tends to zero with $p$, and in fact $p\mathcal{E}^2(h_{GPS})/\log \log (p)$ is bounded (proved in \cite{gurdogan2021}).
In \cite{goldberg2018} it is conjectured, with numerical support, that $E[p \mathcal{E}^2(h_{GPS})]$ is bounded in $p$, and hence the expected variance forecast ratio remains bounded.
Moreover, asymptotically $h_{GPS}$ is closer than $h_{PCA}$ to the true value $b$ in the $\ell_2$ norm; and it yields a portfolio with better tracking error against the true minimum variance portfolio.

\subsection{Our contributions}

The purpose of this paper is to generalize the GPS estimator by introducing a way to use additional information about beta to adjust the shrinkage target $q$ in order to improve the estimate. 

We can consider the space of all possible shrinkage targets $\tau$ as determined by the family of all nontrivial proper linear subspaces $L$ of $\bR^p$ as follows. Given $L$ (assumed not orthogonal to $h$), let the unit vector $\tau(L)$ be the normalized projection of $h$ onto $L$. $\tau(L)$ is then a shrinkage target for $h$ determined by $L$ (and $h$).
We will describe such a subspace $L$ as the linear span of a set of unit vectors called ``anchor points".  In the case of a single anchor point $q$, note that $\tau(\text{span}\{q\}) = q$,  so this case corresponds to the GPS shrinkage target.

The ``MAPS" estimator is a shrinkage estimator with a shrinkage target defined by an arbitrary collection of anchor points, usually including $q$.  When $q$ is the only anchor point, the MAPS estimator reduces to the GPS estimator.
We can therefore think of the MAPS approach as allowing for the incorporation of additional anchor points when this provides additional information.

In Theorem \ref{T1}, we show that expanding $\text{span}\{q\}$ by adding additional anchor points at random asymptotically does no harm, but makes no improvement.

In Theorem \ref{T3}, we show that if the user has certain mild {\it a priori} rank ordering information about groups of components of $\beta$, even with no information about magnitudes, an appropriately constructed MAPS estimator converges exactly to the true vector $b$ in the asymptotic limit.

Theorem \ref{T4} shows that if the betas have positive serial correlation over recent history, then adding the prior PCA estimator $h$ as an anchor point improves the $\ell_2$ error in comparison with the GPS estimator, even if the GPS estimator is computed with the same total data history.

The benefit of improving the $\ell_2$ error in addition to the optimization bias is that it also allows us to reduce the tracking error of the estimated minimum variance fully invested portfolio, discussed in Section \ref{sec:trackingerror} and Theorem \ref{thm:trackingerror}.

In the next sections we present the main results. The framework, assumptions, and statements of the main theorems are presented in Sections \ref{sec:maintheorems} and \ref{sec:trackingerror}. Some simulation experiments are presented in Section \ref{sec:simulations} to illustrate the impact of the main results for some specific situations.

Proofs of the theorems of Section \ref{sec:maintheorems} are organized in   Section \ref{sec:oracletheorems}, with the proofs of some of the needed technical propositions and lemmas appearing in Section \ref{sec:supplement}. Additional details and computations may be found in \cite{gurdogan2021}.


%% file: maps-maintheorems.tex

\subsection{Assumptions and Definitions}

We consider a simple random sample history generated from the basic model (\ref{eq:model}).
The sample data can be summarized as
\begin{equation} \label{eq:onefactormodel}
    R=\beta X^{T}+Z
\end{equation}
where $R\in \mathbb{R}^{p\times n}$ holds the observed individual (excess) returns of $p$ assets for a time window that is set by $n$ consecutive observations. We may consider the observables $R$  to be generated by non-observable random variables $\beta\in \mathbb{R}^{p}$, $X\in \mathbb{R}^{n}$ and $Z\in \mathbb{R}^{p\times n}$. 

The entries of $X$ are the market factor returns for each observation time; the entries of $Z$ are  the specific returns for each asset at each time; the entries of $\beta$ are the exposure of each asset to the market factor, and we interpret $\beta$ as random but fixed at the start of the observation window of times $1,2,3,...,n$ and remaining constant throughout the window.  Only $R$ is observable.
\medskip

In this paper we are interested in asymptotic results as $p$ tends to infinity with $n$ fixed.  Therefore we consider
equation (\ref{eq:onefactormodel})
as  defining an infinite sequence of models, one for each $p$.

To specify the relationship between models with different values of $p$, we need a  more precise  notation.  We'll let $\beta$ refer to an infinite sequence $(\beta(1), \beta(2), \dots ) \in \mathbb{R}^{\infty}$, and 
$\beta^{p} = (\beta(1), \dots, \beta(p)) \in \mathbb{R}^p$ the vector obtained by truncation after $p$ entries.  When the value $p$ is understood or implied, we will frequently drop the superscript and write $\beta$ for $\beta^p$.

Similarly, $Z \in \mathbb{R}^{\infty \times n}$ is a vector of $n$ sequences (the columns), and
$Z^p \in \mathbb{R}^{p \times n}$ is obtained by truncating the sequences at $p$.

With this setup, passing from $p$ to $p+1$ amounts to simply adding an additional asset to the model without changing the existing $p$ assets. The $p$th model is denoted
\[
R^p = \beta^p X^T + Z^p,
\]
but for convenience we will often drop the superscript $p$ in our notation when there is no ambiguity, in favor of equation (\ref{eq:onefactormodel}).

Let $\mu_p(\beta)$ and $d_p(\beta)$ denote the mean and dispersion of $\beta^p$, given by
\begin{equation}
    \mu_p(\beta)=\frac{1}{p}\sum\limits_{i=1}^{p}\beta(i) \text{ }\text{ and }\text{ } d_p(\beta)^2=\frac{1}{p}\sum\limits_{i=1}^{p}(\frac{\beta(i)-\mu_p(\beta)}{\mu_p(\beta)})^2.
\end{equation}
We  make the following assumptions regarding $\beta$, $X$ and $Z$:
\begin{enumerate}
    \item[A1.] (Regularity of beta) The entries $\beta(i)$ of $\beta$ are uniformly bounded, independent random variables, fixed at time 1. The mean $\mu_p(\beta)$ and dispersion  $d_p(\beta)$ converge to limits $\mu_{\infty}(\beta)\in(0,\infty)$ and $d_{\infty}(\beta)\in (0,\infty)$. 
    \item[A2.] (Independence of beta, X, Z) $\beta$, $X$ and $Z$ are jointly independent of each other.

    \item[A3.] (Regularity of X) The entries $X_i$ of $X$ are iid random variables with mean zero, variance $\sigma^2$ \label{a3}. 
    \item[A4.] (Regularity of Z) The entries $Z_{ij}$ of $Z$ have mean zero, finite variance $\delta^2$, and uniformly bounded fourth moment.  In addition,
    the $n$-dimensional rows of $Z$ are mutually independent, and within each row the entries are pairwise uncorrelated.\footnote{Note we do not assume $\beta, X$, or $Z$ are Normal or belong to any specific family of distributions.}

     \label{a4} 
\end{enumerate}

\bigskip

We carry out our analysis with the projection of the vectors on the unit sphere $\mathbb{S}^{p-1} \subset \bR^p$. To that end we define
\begin{equation}
    b=\frac{\beta}{||\beta||} \text{ , } q=\frac{e}{\sqrt{p}},
\end{equation}
where $e = e^p = (1,1, \dots, 1) \in \mathbb{R}^p$, and $||.||$ denotes the usual Euclidean norm.
With the given assumptions the covariance matrix $\Sigma_{\beta}$ of $R$, conditional on $\beta$, is
\begin{equation}
    \Sigma_{\beta}=\sigma^2\beta\beta^T+\delta^2 I.
\end{equation}
Since $\beta$ stays constant over the $n$ observations, the sample covariance matrix $\frac{1}{n}RR^T$ converges to $\Sigma_{\beta}$ almost surely if $n$ is taken to $\infty$, and is the maximum likelihood estimator of $\Sigma_{\beta}$. 


 Since $b$ is a leading eigenvector of $\Sigma_{\beta}$ (corresponding to the largest eigenvalue), then the PCA estimator $h$ (the unit leading eigenvector $h$ of the sample covariance matrix $\frac{1}{n}RR^T$) is a natural estimator of $b$. (We always select the choice of unit eigenvector $h$ such that $(h,q) \geq 0$.)

Since $\beta$ and $X$ only appear in the model $R = \beta X + Z$ as a product, there is a scale ambiguity that we can resolve by combining their scales into a single parameter $eta$:  

\[
\eta^p = \frac{1}{p} |\beta^p|^2 \sigma^2.
\]
It is easy to verify that 
\[
\eta^p = \mu_p(\beta)^2(d_p(\beta)^2+1) \sigma^2,
\]
and therefore by our assumptions $\eta^p$ tends to a positive, finite limit $\eta^{\infty}$ as $p \to \infty$.

Our covariance matrix becomes
\begin{equation}
\Sigma_{\beta} \equiv \Sigma_b = p \eta bb^T + \delta^2 I,
\end{equation}
where we drop the superscript $p$ when convenient.
The scalars $\eta, \delta$ and the unit vector $b$ are to be estimated by $\hat{\eta}$, $\hat{\delta}$, and $h$. As described above, asymptotically only the estimate $h$ of $b$ will be significant.  Improving this estimate is the main technical goal of this paper.

In \cite{goldberg2018} the PCA estimate $h$ is replaced by an estimate $h_{GPS}$ that is ``data driven", meaning that it is computable solely from the observed data $R$. We henceforth use the notation $h_{GPS} = \hat{h}_q$, for a reason that will be clear shortly.  As an intermediate step
we also consider a non-observable ``oracle" version $h_q$, defined as the orthogonal projection in $\mathbb{S}^{p-1}$ of $b$ onto the geodesic joining $h$ to $q$. The oracle version is not data driven because it requires knowledge of the unobserved vector $b$ that we are trying to estimate, but it is a useful concept in the definition and analysis of the data driven version.
Both the data driven estimate $\hat h_q$ and the oracle estimate $h_q$ can be thought of as
obtained from the eigenvector $h$ via ``shrinkage" along the geodesic connecting $h$ to the anchor point, $q$.

 The GPS data-driven estimator $\hat{h}_q$ is successful in improving the variance forecast ratio, and in arriving at a better estimate of the true variance of the minimum variance portfolio. In this paper we have the additional goal of reducing
 the $l_2$ error of the estimator, which, for example, is helpful in reducing tracking error. To that end, we introduce the following new data driven estimator, denoted $\hat{h}_L$.

Let $L_p \subset \mathbb{R}^p$ denote a nontrivial proper linear subspace of $\mathbb{R}^p$. We will sometimes drop the dimension $p$ from the notation.  Denote by $k_p$ the dimension of
$L_p$, with $1 \leq k_p \leq p-1$.   

Let $h^p$ denote the normalized leading eigenvector
of $\frac{1}{n}R^p(R^p)^T$, $s_p^2$ its largest eigenvalue, and $l_p^2$ the average of the remaining eigenvalues. Then we define the data driven ``MAPS"  (Multi Anchor Point Shrinkage) estimator by
\begin{equation} 
    \hat{h}_L=\frac{\tau_ph+\underset{L}{\operatorname{proj}}(h)}{||\tau_ph+\underset{L}{\operatorname{proj}}(h)||} \text{ }\text{ where }\text{ } \tau_p=\frac{\psi_p^2-||\underset{L}{\operatorname{proj}}(h)||^2}{1-\psi_p^2} \label{data}
\end{equation}

and
\begin{equation}
    \psi_p=\sqrt{\frac{s_p^2-l_p^2}{s_p^2}}
\end{equation}
is the relative gap between $s_p^2$ and $l_p^2$. 

\begin{lem}[\cite{goldberg2018}]\label{lem:psi-infty}
The limits 
\[
\psi_{\infty} = \lim_{p \to \infty} \psi_p \text{ and } (h,b)_{\infty} = \lim_{p \to \infty} (h^p,b^p)
\]
 exist almost surely, and
\[
\psi_{\infty} = (h,b)_{\infty} \in (0,1).
\]
\end{lem}

When $L$ is the one-dimensional subspace spanned by the vector $q$, then $\hat{h}_L$ is precisely the GPS estimator $\hat{h}_q$, located along the spherical geodesic connecting $h$ to $q$.   The phrase ``multi anchor point" comes from thinking of $q$ as an ``anchor point" shrinkage target in the GPS paper, and $L$ as a subspace spanned one or more anchor points.  The new shrinkage target determined by $L$ is the normalized orthogonal projection of $h$ onto $L$. When $L$ is the one-dimensional subspace spanned by $q$, the normalized projection of $h$ onto $L$ is just $q$ itself. In the event that $L$ is orthogonal to $h$, the MAPS estimator $\hat h_L$ reverts to $h$ itself.


\subsection{The MAPS estimator with random extra anchor points}

Does adding anchor points to create a MAPS estimator from a higher-dimensional subspace improve the estimation?  The answer depends on whether there is any relevant information in the added anchor points.

We need the concept of a {\em random linear subspace} of $\mathbb{R}^p$. Let $k_p$ be a positive integer such that $1 \leq k_p \leq p-1$. Let
$\xi^p$ be an $O(p)$-valued random variable, where $O(p)$ denotes the orthogonal group in $\mathbb{R}^p$.  Let $\{e^p_1, e^p_2, \dots, e^p_p \}$ denote the standard Cartesian basis of $\bR^p$.

We say that $L_p$ is a {\em random linear subspace} of $\bR^p$ with dimension $k_p$ if, for some $\xi^p$ as above,
\[
L_p = \text{span}_p\{\xi^p e^p_i | i = 1,2, \dots, k_p \},
\]
where $\text{span}_p$ denotes the linear span of a set of vectors in $\bR^p$.

We say $L_p$ is independent of a random variable $X$ if the generator $\xi^p$ is independent of $X$.
Moreover, we say $H_p$ is a {\em Haar random subspace} of $\bR^p$ if it is a random linear subspace as above, and the random variable $\xi^p$ induces the (uniform) Haar measure on $O(p)$.

\begin{defn}
A non-decreasing sequence $\{k_p \}$ of positive integers is {\em square root dominated}
if  
\[
\sum_{p=1}^{\infty} \frac{k_p^2}{p^2} < \infty .
\]
\end{defn}
For example, any non-decreasing sequence satisfying $k_p \leq C p^{\alpha}$ for $\alpha < 1/2$ is square root dominated.

\begin{thm}\label{T1}
Let the assumptions \textbf{1},\textbf{2},\textbf{3} and \textbf{4} hold. Suppose, for each $p$,  $L_p$ is a random linear subspace and $H_p$ is a Haar random subspace of $\bR^p$. Suppose also that $L_p$ and $H_p$ are independent of $\beta$ and $Z$, and the sequences $\dim L_p$ and $\dim H_p$ are square root dominated.  

Let $L'_p = \text{span}\{L_p, q^p\}$ and $L''_p = \text{span}\{H_p, q^p\}$.

Then, almost surely,
\begin{eqnarray}
(a) & \limsup\limits_{p\rightarrow \infty}||\hat{h}_{L'}-b|| &\leq ||\hat{h}_q-b||_{\infty}, \\
(b) &  \lim\limits_{p\rightarrow \infty}||\hat{h}_{L''}-b|| &=||\hat{h}_q-b||_{\infty}, \text{ and } \label{eq:Haar+q} \\
(c) &  \lim\limits_{p\rightarrow \infty}||\hat{h}_H-b|| &=||h-b||_{\infty}.
\end{eqnarray}

\end{thm}


Theorem \ref{T1} says adding random anchor points to form a MAPS estimator does no harm, but also makes no improvement asymptotically. Equation (\ref{eq:Haar+q}) says that the GPS estimator is neither improved nor harmed by adding extra anchor points uniformly at random. Therefore the goal will be to find useful anchor points that take advantage of additional information that might be available.

\subsection{The MAPS estimator with rank order information about the entries of beta}

As with stocks grouped by sector, it may be that the betas can be separated into ordered groups, where the rank ordering of the groups is known, but not the ordering within groups.
This turns out to be enough information for the MAPS estimator to converge asymptotically to the true value almost surely.

\begin{defn}
For any $p\in\mathbb{N}$, let $\mathcal{P}=\mathcal{P}(p)$ be a partition of the index set $\{1,2,..,p\}$ (i.e.~a collection of pairwise disjoint non-empty subsets, called atoms, whose union is $\{1,2,..,p\}$). The number of atoms of $\mathcal{P}$ is denoted by $|\mathcal{P}|$. 

We say the sequence of partitions $\mathcal{P}(p)$ is {\bf semi-uniform} if there exists $M>0$ such that for all $p$,
 \begin{equation}
      \max\limits_{I\in \mathcal{P}(p)}|I|\leq  M\frac{p}{|\mathcal{P}(p)|}  .
    \end{equation}
In other words, no atom is larger than a multiple $M$ of the average atom size.

Given $\beta \in \bR^p$, we say $\mathcal{P}$ is $\beta$-\textbf{ordered} if, for each distinct $I,J\in \mathcal{P}$, either $\max\limits_{i\in I}\beta_i\leq \min\limits_{j\in J}\beta_j$ or $\max\limits_{j\in J}\beta_j\leq \min\limits_{j\in I}\beta_i$.
\end{defn}

\begin{defn}
For any $A \subset \{1,2,...,p\}$ define a unit vector $v^A\in \mathbb{R}^p$ by
\begin{equation}
    v^A(i)=1_A(i)\frac{1}{\sqrt{|A|}},
\end{equation}
where $1_A$ denotes the indicator function of $A$.
We may then define, for any partition $\mathcal{P}=\mathcal{P}(p)$, an induced linear subspace $L(\mathcal{P})$ of $\mathbb{R}^p$ by
\begin{equation}
    L(\mathcal{P})=\text{span}_p\{v^A\big| A\in \mathcal{P}\} \equiv < v^A \big| A \in \mathcal{P} >.
\end{equation}
\end{defn}
\begin{thm}\label{T3}
Let the assumptions \textbf{1},\textbf{2},\textbf{3} and \textbf{4} hold. 
Consider a semi-uniform sequence  $\{\mathcal{P}(p) : p=1,2,3,\dots \}$ of $\beta$-ordered partitions
such that the sequence $\{ |\mathcal{P}(p)| \}$ tends to infinity and is square root dominated. Then 
\begin{equation}
    \lim\limits_{p \rightarrow \infty}||\hat{h}_{L(\mathcal{P}(p))}-b||=0 \text{ }\text{ almost surely. }
\end{equation}

\end{thm}

  Theorem \ref{T3} says that if we have certain prior information about the ordering of the $\beta$ elements in the sense of finding an ordered partition (but with no prior information about the magnitudes of the elements or their ordering within partition atoms), then asymptotically we can estimate $b$ exactly.

Having in hand a genuine ordered partition {\it a priori} is likely only approximately possible
in the real world.   Theorem \ref{T3} is suggestive that even partial grouped order information about the betas can  be helpful in strictly improving the GPS estimate. This is confirmed empirically in section \ref{sec:simulations}.   

The next theorem shows that even with no {\it a priori} information beyond the observed time series of returns, we can still use MAPS to improve the GPS estimator.

\subsection{A data-driven dynamic MAPS estimator}

In the analysis above we have treated $\beta$ as a constant throughout the sampling period, but in reality we expect $\beta$ to vary slowly over time. To capture this in a simple way, let's now assume that we have access to returns observations for $p$ assets over a fixed number of $2n$ periods. The first $n$ periods we call the first (or previous) time block, and the second $n$ periods the second (or current) time block.  We then have returns matrices $R_1, R_2 \in \bR^{p \times n}$ corresponding to the two time blocks, and $R = [R_1 R_2] \in \bR^{p \times 2n}$ the full returns matrix over the full set of $2n$ observation times.

Define the sample covariance matrices $S, S_1, S_2$ as $\frac{1}{2n}RR^T$, $\frac{1}{n}R_1R_1^T$, and $\frac{1}{n}R_1R_1^T$, respectively. Let $h, h_1, h_2$ denote the respective (normalized) leading eigenvectors (PCA estimators) of $S, S_1, S_2$. (Of the two choices of eigenvector, we always select the one having non-negative inner product with $q$.)

Instead of a single $\beta$ for the entire observation period, we suppose there are random vectors $\beta_1$ and $\beta_2$ that enter the model during the first and second time blocks, respectively, and are fixed during their respective blocks. We assume both $\beta_1$ and $\beta_2$ satisfy assumptions (1) and (2) above, and denote by $b_1$ and $b_2$ the corresponding normalized vectors.  The vectors $\beta_1$ and $\beta_2$ should not be too dissimilar in the mild sense that $(\beta_1, \beta_2) \geq 0$.

\begin{defn}
Define the co-dispersion $d_p(\beta_1,\beta_2)$ and pointwise correlation $\rho_p(\beta_1,\beta_2)$  of $\beta_1$ and $\beta_2$  by
$$d_p(\beta_1,\beta_2)=\frac{1}{p}\sum\limits_{i=1}^{p}\big(\frac{\beta_1(i)}{\mu_{p}(\beta_1)}-1\big)\big(\frac{\beta_2(i)}{\mu_{p}(\beta_2)}-1\big)$$ and
$$\rho_p(\beta_1,\beta_2)=\frac{d_p(\beta_1,\beta_2)}{d_p(\beta_1)d_p(\beta_2)}.$$
\end{defn}

The Cauchy-Schwartz inequality shows $-1 \leq \rho_p(\beta_1,\beta_2) \leq 1$. Furthermore, it is straightforward to verify that
\begin{equation}
    (b_1,b_2) - (b_1,q)(b_2,q) = \frac{d_p(\beta_1, \beta_2)}{\sqrt{1+d_p(\beta_1)^2}\sqrt{1+d_p(\beta_2)^2}}.
\end{equation}

and hence $d_p(\beta_1, \beta_2)$, and $\rho_p(\beta_1, \beta_2)$ have limits $d_{\infty}(\beta_1, \beta_2)$, and $\rho_{\infty}(\beta_1, \beta_2)$ as $p \to \infty$.
\smallskip

Our motivation for this model is the intuition that the betas for different time periods are noisy representations of a fundamental beta, and that the beta from a recent time block provides some useful information about beta in the current time block.  To make this precise in support of the following theorem, we make the following additional assumptions.

\smallskip
\noindent
\begin{enumerate}
\item[A5.] [Relation between $\beta_1$ and $\beta_2$] Almost surely, $(\beta_1, \beta_2) > 0$,  $\mu_{\infty}(\beta_1) = \mu_{\infty}(\beta_2)$, $d_{\infty}(\beta_1) = d_{\infty}(\beta_2)$, and $\lim_{p \to \infty} d_p(\beta_1,\beta_2) = d_{\infty}(\beta_1, \beta_2)$ exists.
\end{enumerate}

\begin{thm} \label{T4}
Assume $\beta_1$, $\beta_2$, $R, X, Z$ satisfy assumptions 1-5. Denote by $\hat{h}_q^s$ and $\hat{h}_q^d$ the GPS estimators for $R_2$ and $R$, respectively, i.e. the current (single) and previous plus current (double) time blocks. Let $h_1$ and $h_2$ be the PCA estimators for $R_1$ and $R_2$, respectively.

Let $L_p = < h_1, q >$ and define a MAPS estimator for the current time block as
\begin{equation} \label{eq:dynamicalmapsestimator}
    \hat{h}_L=\frac{\tau_p h_2+\underset{L}{\operatorname{proj}}(h_2)}{||\tau_p h_2+\underset{L}{\operatorname{proj}}(h_2)||} \text{ }\text{ where }\text{ } \tau_p=\frac{\psi_p^2-||\underset{L}{\operatorname{proj}}(h_2)||^2}{1-\psi_p^2}.
\end{equation}
Then, almost surely,
\begin{enumerate}
\item[(a)]
\begin{equation}
    \lim\limits_{p\rightarrow \infty}\big(||\hat{h}_L-b_2||- ||\hat{h}_q^s-b_2||]\big)\leq 0 \text{ }\text{ and }\text{ } \lim\limits_{p\rightarrow \infty}\big(||\hat{h}_L-b_2||- ||\hat{h}_q^d-b_2||]\big)\leq 0 . \label{a1}
\end{equation}
\item[(b)]
If $0<|\rho_{\infty}(\beta_1,\beta_2)|<1$ almost surely,
\begin{equation}
    \lim\limits_{p\rightarrow \infty}\big(||\hat{h}_L-b_2||-||\hat{h}^s_q-b_2||\big)<0 
    \text{ and }\text{ } \lim\limits_{p\rightarrow \infty}\big(||\hat{h}_L-b_2||-||\hat{h}_q^d-b_2||\big)<0 .\label{a2}
\end{equation}
\end{enumerate}
\end{thm} 

Theorem \ref{T4} says that the MAPS estimator obtained by adding the PCA estimator $h$ from the previous time block as a second anchor point outperforms the GPS estimator asymptotically, as measured by $l_2$ error, even if the latter estimated with the full $2n$ (double) data set. This works when the previous time block carries some information about the current beta (non-zero correlation).  In the case of perfect correlation $\rho_{\infty}(\beta_1,\beta_2)=1$ the two betas are equal, and we then return to the GPS setting where beta is assumed constant, so no improved performance is expected.

 The cost of implementing this ``dynamic MAPS" estimator is comparable to that of the GPS estimator, so should generally be preferred when no rank order information is available for beta.

%% file: maps-trackingerror.tex
Our task has been to estimate the covariance matrix of returns for a large number $p$ of assets but a short time series of $n$ returns observations.

Recall that for the returns model (\ref{eq:model}),
under the given assumptions, we have the true covariance matrix
\[
\Sigma_b = p \eta bb^T + \delta^2 I,
\]
where $\eta$ and $\delta$ are positive constants and $b$ is a unit $p$-vector, and
we are interested in corresponding estimates $\hat \eta$, $\hat \delta$, and $h$ that define an estimator
\[
\Sigma_h = p \hat \eta hh^T + \hat \delta^2 I.
\]

The theorems above are about finding an estimator $h$ of $b$ that asymptotically controls the $\ell_2$ error $|| h - b ||$.
We are ignoring $\hat \eta$ and $\hat \delta$ because of Proposition \ref{prop:gpsvariance}, showing that the true variance of the estimated minimum variance portfolio $\hat w$, and the variance forecast ratio, are asymptotically controlled by $h$ via the optimization bias
\[
\mathcal{E}(h) = \frac{(b,q) - (b,h)(h,q)}{1-(h,q)^2}.
\]

We now turn to another important measure of portfolio estimation quality: the tracking error.


Recall that $w$ denotes the true minimum variance portfolio using $\Sigma$, and $\hat w$ is the minimum variance portfolio using the estimated covariance matrix $\hat \Sigma$.

\begin{defn}
The (true) tracking error $\mathcal{T}(h)$ associated to $\hat w$ is defined by
\begin{equation}
\mathcal{T}^2(h) = (\hat w - w)^T \Sigma (\hat w - w).
\end{equation}
\end{defn}

\begin{defn}
Given the notation above, define the {\em eigenvector bias} $\mathcal{D}(h)$ associated to a unit leading eigenvector estimate $h$  as
\[
\mathcal{D}(h) = \frac{(h,q)^2(1-(h,b)^2)}{(1-(h,q)^2)(1-(b,q)^2)} = \frac{(h,q)^2||h-b||^2}{||h-q||^2||b-q||^2}.
\]

\end{defn}

\begin{thm} \label{thm:trackingerror}
Let $h$ be an estimator of $b$ such that $\mathcal{E}(h) \to 0$ as $p \to \infty$ (such as a GPS or MAPS estimator).
Then the tracking error of $h$ is asymptotically (neglecting terms of higher order in $1/p$) given by
\begin{equation} \label{eq:trackingerror}
\mathcal{T}^2(h) = \eta \mathcal{E}^2(h) + \frac{\delta^2}{p} \mathcal{D}(h) +  \frac{C}{p}\mathcal{E}(h),
\end{equation}
where
\[
C = \frac{2}{\xi (1+d_{\infty}^2(\beta))} (\delta^2 + \frac{\eta}{\hat \eta} \hat \delta^2)
\]
and
$\xi > 0$ is a constant depending only on $\psi_{\infty}$, $\mu_{\infty}(\beta)$, and $d_{\infty}(\beta)$. 
\end{thm} 

We consider what this theorem means for various estimators $h$.  For the PCA estimate,
it was already shown in \cite{goldberg2018} that $\mathcal{E}(h_{PCA})$ is asymptotically bounded below, and hence so
is the tracking error.

On the other hand, $\mathcal{E}(h_{GPS})$ tends to zero as $p \to \infty$.  In addition
\cite{goldberg2018} shows that
\[
\limsup_{p \to \infty} p \, \mathcal{E}^2(h_{GPS}) = \infty
\]
almost surely, while  \cite{gurdogan2021} shows
\[
\limsup_{p \to \infty} \frac{p \,\mathcal{E}^2(h_{GPS})}{ \log \log p} < \infty,
\]
and we conjecture the same is true for the more general estimator $h_{MAPS}$.

This implies the leading terms, asymptotically, are
\[
\mathcal{T}^2(h_{MAPS}) \leq \eta \mathcal{E}^2(h_{MAPS}) + (\delta^2/p) \mathcal{D}(h_{MAPS})
\]

Note here the estimated parameters $\hat \eta$ and $\hat \delta$ have dropped out, with the tracking error asymptotically controlled by the eigenvector estimate $h$ alone.

Theorem \ref{thm:trackingerror} helps justify our interest in the $\ell_2$ error results of Theorems \ref{T3} and \ref{T4}.  Reducing the $\ell_2$ error $||h-b||$ of the $h$ estimate controls the second term $\mathcal{D}(h)$ of the asymptotic estimate for tracking error.  We therefore expect to see improved total tracking error when we are able to make an informed choice of additional anchor points in forming the MAPS estimator.  This is borne out in our numerical experiments described in Section \ref{sec:simulations}. 

\bigskip
\noindent
{\em Proof of Theorem \ref{thm:trackingerror}}

\begin{lem} \label{lemma:trackingerror1}
There exists $\xi > 0$, depending only on $\psi_{\infty}$, $\mu_{\infty}(\beta)$, and $d_{\infty}(\beta)$,
such that for any $p$ sufficiently large, and any linear subspace $L$ of $\bR^p$ that contains $q$,
\[
||h_{L} - q ||^2 > \xi > 0,
\]
where $h_{L}$ is the MAPS estimator determined by $L$.

\end{lem}

The Lemma follows from the fact that $(h_L,q) \leq (h_{GPS},q)$, and is proved for the case $h_{GPS}$ using the definitions and the known limits
\begin{eqnarray}
(h_{PCA}, q)_{\infty} &=& (b,q)_{\infty} (h_{PCA}, b)_\infty \\
(b,q)^2_{\infty} &=& \frac{1}{1+d_{\infty}^2(\beta)} \in (0,1) \label{eq:(b,q)}\\
(h_{PCA}, b)_{\infty} &=& \psi_{\infty} > 0.
\end{eqnarray}

From the Lemma and equation (\ref{eq:(b,q)}), we may assume without loss of generality that
 $\xi > 0$ is an asymptotic lower bound for both $||h_L - q||^2 = 1 - (h_L,q)^2$ and $||b -q||^2 = 1 - (b,q)^2$.

Next, we recall it is straightforward to find explicit formulas for the minimum variance portfolios $w$ and $\hat w$:
\[
w = \frac{1}{\sqrt{p}} \frac{\rho q - b}{\rho - (b,q)}, \quad \text{ where } \rho = \frac{1+k^2}{(b,q)}, \quad k^2 = \frac{\delta^2}{p\eta}
\]
and
\[
\hat w = \frac{1}{\sqrt{p}} \frac{\hat \rho q - h}{\hat \rho - (h,q)}, \quad \text{ where } \hat \rho = \frac{1+\hat k^2}{(h,q)}, \quad \hat k^2 = \frac{\hat \delta^2}{p\hat\eta}.
\]

We may use these expressions to obtain an explicit formula for the tracking error:

\begin{eqnarray*}
\mathcal{T}^2(h) &=& (\hat w - w)^T \Sigma (\hat w - w) = (\hat w - w)^T (p\eta bb^T + \delta^2I) (\hat w - w) \\
&=& p\eta(\hat w - w, b)^2 + \delta^2||\hat w - w||^2.
\end{eqnarray*}

We now estimate the two terms on the right hand side separately.

(1) For the first term $p\eta(\hat w - w, b)^2$, it is convenient to introduce the notation
\[
\Gamma = \frac{k^2}{1+k^2-(b,q)^2} \text{ and } \hat \Gamma = \frac{\hat k^2}{1+\hat k^2-(h,q)^2},
\]
and since 
\[
\Gamma \leq \frac{k^2}{\xi} \text{ and } \hat \Gamma \leq \frac{\hat k^2}{\xi}
\]
both $\Gamma$ and $\hat \Gamma$ are of order $1/p$.

A straightforward computation verifies that
\begin{eqnarray}
(w,b) &=& \frac{1}{\sqrt{p}} \Gamma (b,q) \\
(\hat w, b) &=& \frac{1}{\sqrt{p}} \left( \mathcal{E}(h) + \hat \Gamma [ (b,q) - \mathcal{E}(h) ] \right).
\end{eqnarray}

We then obtain

\begin{eqnarray}
p(\hat w - w, b)^2 &=& p[ (\hat w,b) - (w,b)]^2 \\
&=& \mathcal{E}(h)^2 + 2\mathcal{E}(h) G + G^2,
\end{eqnarray}

where $G =  \hat \Gamma ((b,q) - \mathcal{E}(h)) - \Gamma (b,q)$.

Since asymptotically $(b,q)$ is bounded below and $\mathcal{E}(h) \to 0$, the third term $G^2$ is of order $1/p^2$ and can be dropped.  We thus obtain the asymptotic estimate
\[
p (\hat w - w,b)^2 \leq \mathcal{E}^2 + 2 \mathcal{E}(h)(\hat \Gamma - \Gamma)(b,q).
\]

Multiplying by $\eta$ and using the bounds on $\Gamma, \hat \Gamma$ and the limit of $(b,q)$, we obtain
\[
p \eta (\hat w - w,b)^2 \leq \mathcal{E}^2 + \frac{C}{p} \mathcal{E}(h),
\]
where $C$ is the constant defined in the statement of the theorem.

\smallskip

(2) We now turn to the second term $||\hat w - w||^2 = ||\hat w||^2 + ||w||^2 - 2(\hat w, w)$.

Using the definitions of $\hat w$ and $w$ and the fact that $k^2$, $\hat k^2$ are of order $1/p$, after a calculation we obtain, to lowest order in $1/p$,
\begin{equation}
p ||\hat w - w||^2 = \frac{(h,q)^2[1-(h,b)^2]}{(1-(h,q)^2)(1-(b,q)^2)} + \frac{1-(h,q)^2}{1-(b,q)^2} \mathcal{E}^2(h).
\end{equation}

Since $\mathcal{E}(h) \to 0$, we may neglect the second term, and putting (1) and (2) together yields
\[
\mathcal{T}^2(h) \leq  \mathcal{E}^2 + \frac{C}{p} \mathcal{E}(h) + \frac{\delta^2}{p} \mathcal{D}(h).
\]

%% file: maps-simulations.tex
To illustrate the previous theorems, we present the results of numerical experiments showing the improvement that MAPS estimators can bring in estimating the covariance matrix. To approximate the asymptotic regime, in these experiments we use $p=500$ stocks. 
The Python code used to run these experiments and create the figures is available at \\ \texttt{https://github.com/hugurdog/MAPS\_NumericalExperiments}.

\subsection{Simulated betas with correlation}

First we set up a test bed consisting of a double block of data where we can control the beta correlation. Set $n = 24$. We generate observations $R_i^1, R_i^2\in \mathbb{R}^p$ for $i=1,2,...,n$, according to the market model of Equation (\ref{eq:model}):
\begin{equation} \label{eq:2blocksimulationmodel}
    R_i^t=\beta_t X_i^t+ Z_{i}^t \text{ }\text{, }t=1,2\text{ }\text{, }i=1,2,...,n
\end{equation}
for unobserved market returns $X^t_i \in \mathbb{R}$ and unobserved asset specific returns $Z^t_i\in \mathbb{R}^p$ for each time window of data.

Here the $p\times n$ matrices $R^1$ and $R^2$ represent the previous and current block of $n$ consecutive excess returns, respectively, and are obtained from Equation (\ref{eq:2blocksimulationmodel}) by randomly generating
 $\beta, X$, and $Z$:
\begin{itemize}
    \item the market returns $X^t_i$ are an iid random sample drawn from a normal distribution with mean 0 and variance $\sigma^2 = 0.16$,
    \item the asset specific returns $\{Z^1_i\}_1^n$ and $\{Z^2_i\}_1^n$ are  i.i.d. normal with mean $0$ and variance $\delta^2I=(.5)^2I$, and
    \item the $p$-vectors $\beta_1$ and $\beta_2$ are drawn independently of $X$ and $Z$ from a normal distribution with mean $0$ and variance $(.5)^2I$ and with pointwise correlation $\rho_p(\beta^1, \beta^2)  \in [0,1]$ for a range of values of $\rho$ specified below.\footnote{exact recipe for $\beta_1, \beta_2$ here.}
\end{itemize}

The true covariance matrix of the $n$ most recent returns $R^2$ is
\begin{equation}
    {\Sigma}={\sigma}^2{\beta_2}{\beta_2}^T+{\delta}^2I,
\end{equation}
which we wish to estimate by
\begin{equation} \label{eq:estimatedcovmatrix}
    \hat{\Sigma}=\hat{\sigma}^2\hat{\beta}\hat{\beta}^T+\hat{\delta}^2I.
\end{equation}
Following the lead of \cite{goldberg2018}, we fix
\begin{equation} \label{eq:estimatedscalars}
    \hat{\sigma}^2|\hat{\beta}|^2=s_p^2-l_p^2 \text{ and } \hat{\delta}^2=\frac{n}{p}l_p^2
\end{equation} 
and vary only the estimator of $ \frac{\hat{\beta}}{|\hat{\beta}|} = h$. In our numerical experiments we compare performance for the following choices of $h$:
\begin{enumerate}
    \item $h^s$ the PCA estimator on the single block $R^2$ (PCA1)
    \item $h^d$ the PCA estimator on the double block $R=[R^1,R^2]$ (PCA2)
    \item $\hat{h}_q^s$, the GPS estimator on the single block $R^2$ (GPS1)
    \item $\hat{h}_{L{_D}}$ the dynamical MAPS estimator defined on the double block of data $R=[R^1,R^2]$ by the equation \textbf{(18)}.(Dynamical MAPS)
    \item $\hat{h}_q^d$, the GPS estimator on the double block $R=[R^1,R^2]$ (GPS2)
    \item $\hat{h}^s_{L(\mathcal{P})}$ is the MAPS estimator on single block $R^2$ where $\mathcal{P}$ is a beta ordered uniform partition constructed by using the ordering of the entries of $\beta^2$ and where the number of atoms $k_p$ in each partition is set to 8, which is approximately $\sqrt[3]{p}$.\footnote{The largest 479 beta values are partitioned into 7 groups of 71, and the three lowest values form the eighth partition atom.} (Beta Ordered MAPS)
\end{enumerate}

We report the performance of each of these estimators according to the following two metrics:
\begin{itemize}
    \item The $l_2$ error $||b - h||$ between the true normalized beta $b=\frac{\beta^2}{|\beta^2|}$ of the current data block and the estimated version $h = \frac{\hat{\beta}}{|\hat{\beta}|}$.
    \item The tracking error between the true and estimated minimum variance portfolios $w$ and $\hat{w}$
\begin{equation}
    \mathcal{T}^2(\hat{w})=(\hat{w}-w)^T\Sigma (\hat{w}-w).
\end{equation}
\end{itemize} 
Results of the comparison are displayed below for  values of the pointwise correlation $\rho$ selected from $\{0,0.2,.4,.6,.8,1\}$. For each choice of $\rho$, the experiment was run 100 times, resulting in 100 $\ell_2$ and tracking error values each.
These values are summarized using standard box-and-whisker plots generated in Python using the package matplotlib.pyplot.boxplot.  

\begin{figure}[!htb]
    \centering
    \includegraphics[width=\textwidth]{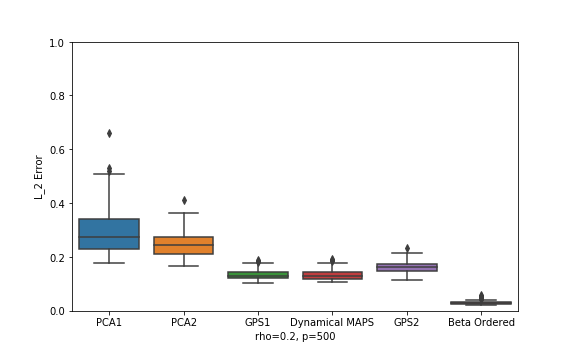}
    \caption{The $\ell_2$ error performance of six different estimators as defined in the text.  Here the pointwise correlation of betas between the two time blocks is $\rho = 0.2$.}
    \label{fig:simulation-L2case}
\end{figure}

Figure \ref{fig:simulation-L2case} shows the $\ell_2$ error $||h-b||$ for different estimators $h$ (in the same order, left to right, as listed above) for the case $\rho = 0.2$.  The worst performer is the single block PCA.  (It is independent of $\rho$ since it doesn't see the earlier data at all.)  Double block PCA is a little better, but the other estimators are far better.
Since GPS effectively assumes that the betas have perfect serial correlation, it's not surprising that the double block GPS does slightly worse than the single block in this case.
The best estimator is the MAPS estimator with prior information about group ordering.  Assuming no such information is available, the GPS2 and Dynamical MAPS estimators are about tied for best.

\begin{figure}[!htb]
    \centering
    \begin{subfigure}[b]{0.48\textwidth}
    \centering
    \includegraphics[width=\textwidth]{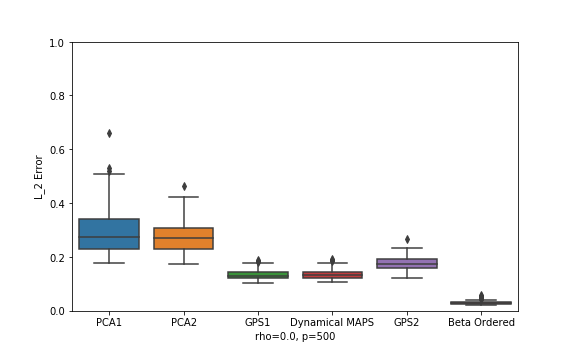}
    \caption{$\rho = 0$}
    \label{fig:l2rho=0}
    \end{subfigure}
    \hfill
    \begin{subfigure}[b]{0.48\textwidth}
    \centering
    \includegraphics[width=\textwidth]{figs/Numerical_Double_Block_L2_Error2_500.png}
    \caption{$\rho = 0.2$}
    \label{fig:l2rho=2}
    \end{subfigure}
    \hfill
    \begin{subfigure}[b]{0.48\textwidth}
    \centering
    \includegraphics[width=\textwidth]{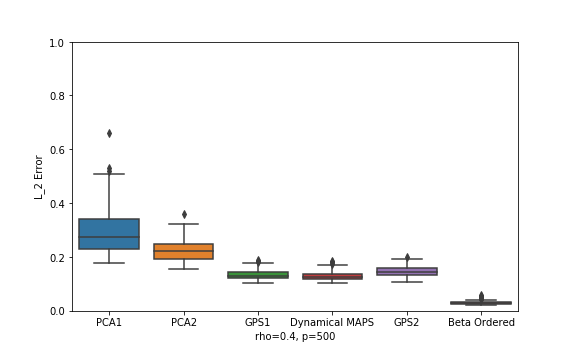}
    \caption{$\rho = 0.4$}
    \label{fig:l2rho4}
    \end{subfigure}
        \hfill
    \begin{subfigure}[b]{0.48\textwidth}
    \centering
    \includegraphics[width=\textwidth]{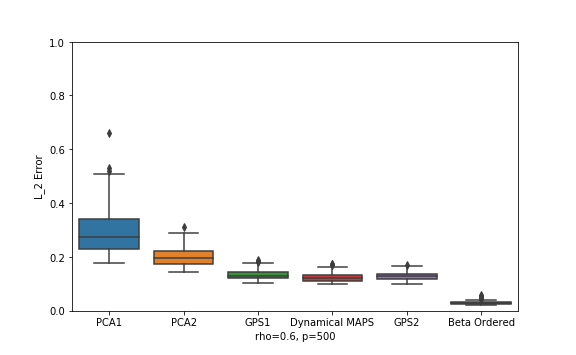}
    \caption{$\rho = 0.6$}
    \label{fig:l2rho6}
    \end{subfigure}
        \hfill
    \begin{subfigure}[b]{0.48\textwidth}
    \centering
    \includegraphics[width=\textwidth]{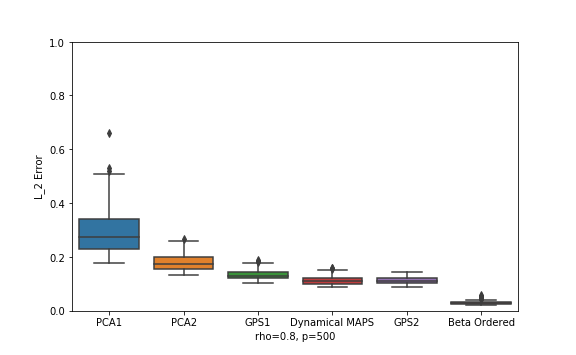}
    \caption{$\rho = 0.8$}
    \label{fig:l2rho8}
    \end{subfigure}
        \hfill
    \begin{subfigure}[b]{0.48\textwidth}
    \centering
    \includegraphics[width=\textwidth]{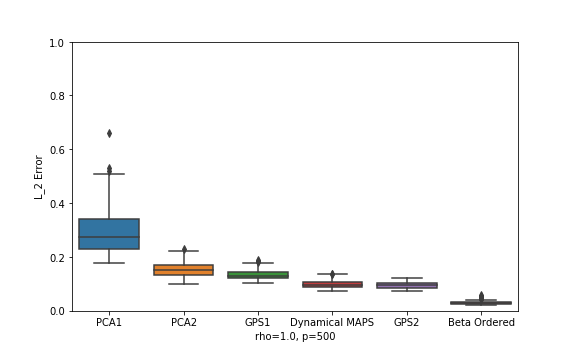}
    \caption{$\rho = 1.0$}
    \label{fig:l2rho10}
    \end{subfigure}
    \caption{Results of simulation experiments for different estimators PCA1, PCA2, GPS1, Random Partition, Dynamical Maps, GPS2, and Beta Ordered.  The pointwise correlation $\rho$ is the correlation between betas in the two different time blocks. Figure \ref{fig:l2rho=2} is the same as Figure \ref{fig:simulation-L2case}.}
    \label{fig:simulation-L2}
    \vfill
\end{figure}

Figure \ref{fig:simulation-L2} shows the results for $\rho = 0, 0.2, 0.4, 0.6, 0.8, 1.0$ in smaller size for visual comparison.   Throughout the range, the dynamical MAPS estimator outperforms all the other purely data-driven estimators, but the beta-ordered MAPS estimator remains in the lead.

\begin{figure}[!htb]
    \centering
    \includegraphics[width=\textwidth]{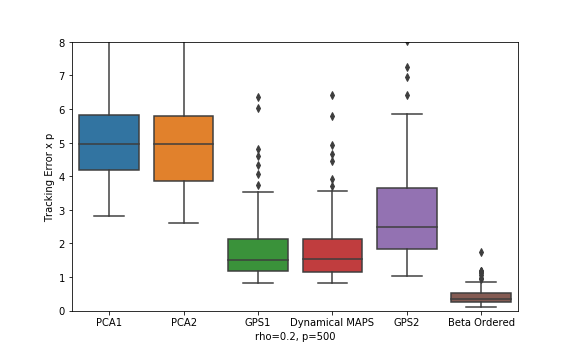}
    \caption{The tracking error error performance of different estimators.  Here the pointwise correlation of betas between the two time blocks is $\rho = 0.2$.}
    \label{fig:simulation-Trackingcase}
\end{figure}

\begin{figure}[!htb]
    \centering
    \begin{subfigure}[b]{0.48\textwidth}
    \centering
    \includegraphics[width=\textwidth]{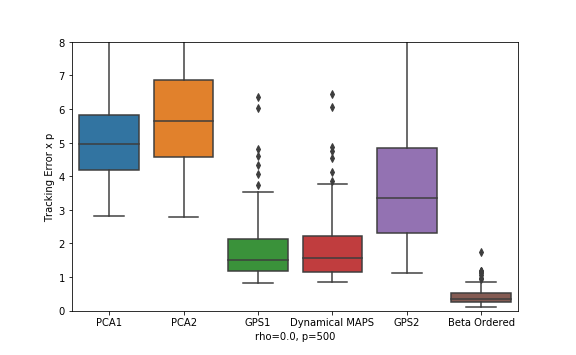}
    \caption{$\rho = 0$}
    \label{fig:Trackingrho=0}
    \end{subfigure}
    \hfill
    \begin{subfigure}[b]{0.48\textwidth}
    \centering
    \includegraphics[width=\textwidth]{figs/Numerical_Double_Block_Tracking_Error2_500.png}
    \caption{$\rho = 0.2$}
    \label{fig:Trackingrho=2}
    \end{subfigure}
    \hfill
    \begin{subfigure}[b]{0.48\textwidth}
    \centering
    \includegraphics[width=\textwidth]{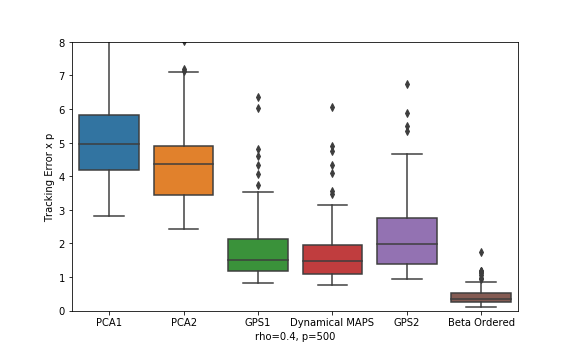}
    \caption{$\rho = 0.4$}
    \label{fig:Trackingrho4}
    \end{subfigure}
        \hfill
    \begin{subfigure}[b]{0.48\textwidth}
    \centering
    \includegraphics[width=\textwidth]{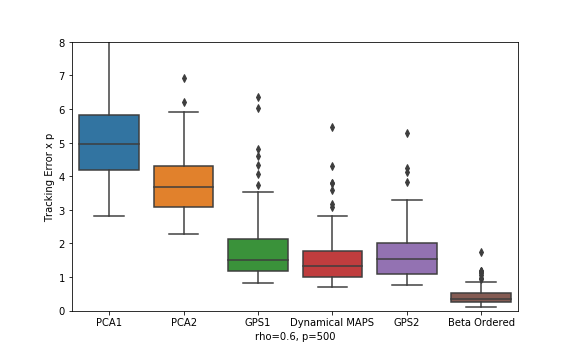}
    \caption{$\rho = 0.6$}
    \label{fig:Trackingrho6}
    \end{subfigure}
        \hfill
    \begin{subfigure}[b]{0.48\textwidth}
    \centering
    \includegraphics[width=\textwidth]{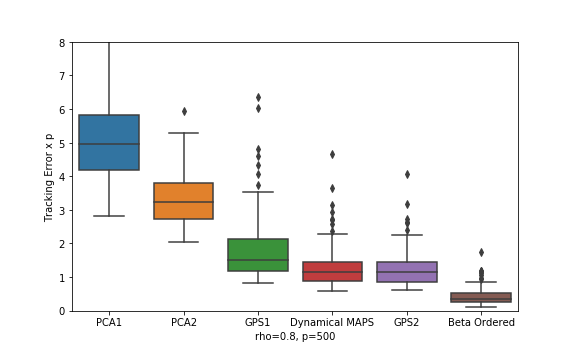}
    \caption{$\rho = 0.8$}
    \label{fig:Trackingrho8}
    \end{subfigure}
        \hfill
    \begin{subfigure}[b]{0.48\textwidth}
    \centering
    \includegraphics[width=\textwidth]{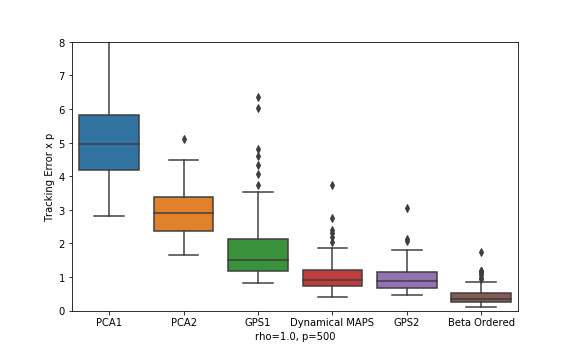}
    \caption{$\rho = 1.0$}
    \label{fig:Trackingrho10}
    \end{subfigure}
    \caption{Tracking error results of simulation experiments for different estimators PCA1, PCA2, GPS1, Random Partition, Dynamical Maps, GPS2, and Beta Ordered.  The pointwise correlation $\rho$ is the correlation between betas in the two different time blocks. Figure \ref{fig:Trackingrho=2} is the same as Figure \ref{fig:simulation-Trackingcase}.}
    \label{fig:simulation-Tracking}
    \vfill
\end{figure}

Figure \ref{fig:simulation-Trackingcase} presents the results for tracking error, reported as $p\mathcal{T}^2$. Results are similar to the $\ell_2$ error, but stronger.  Again, the dynamical MAPS estimator does best among all methods that don't use order information, and the beta ordered MAPS estimator is significantly better than all others.  Figure \ref{fig:simulation-Tracking} displays tracking error outcomes for a range of correlation values $\rho(\beta_1, \beta_2)$.

We conclude from these experiments the dynamical MAPS estimator is best when the only the returns are available, and the beta ordered MAPS estimator is preferred when rank order information on the betas is available.


\newpage


\newpage


\subsection{Simulations with historical betas}

In this section we use historical betas rather than randomly generated ones to test the quality of some MAPS estimators.  We use 24 historical monthly CAPM betas for each of the S\&P 500 firms provided by WRDS\footnote{Wharton Research Data Services, wrds-www.wharton.upenn.edu} between the dates 01/01/2018 and 11/30/2020.  
We denote these betas as $\beta_1, \dots, \beta_{24}$. We will have two different mechanism of generating observations of the market model for the single and double data block test beds. 
\subsubsection{Single Data Block}
The WRDS beta suite estimates beta each month from the prior 12 monthly returns.
Hence we generate $n=12$ sequential observations of the market model for each beta separately,
\begin{equation}
     R_i^t=\beta_t X_i^t+ Z_{i}^t \quad i=1,2,...,12, \quad t=1,2, \dots, 24,
\end{equation}
with the unobserved market return $X^t$ and the asset specific return $Z^t$ generated using the same settings as in the previous section. 

For each $\beta_t$ this produces a $p \times n$ returns matrix $R^t$ from which we can derive the following four estimators $h^t$ of $\beta_t$:

\begin{enumerate}
    \item $h_{PCA}^t$, the PCA estimator of $R^t$. (PCA)
    \item $\hat{h}^t_q$ the GPS estimator of $R^t$. (GPS)
    \item $\hat{h}^t_{L(\mathcal{P}_{s})}$, the MAPS estimator of $R^t$, where $\mathcal{P}_s$ is a sector partitioning of the indices $\{1,2,...,p\}$ in which each atom in the partition contains the indices of one of the 11 sectors in the 
    market\footnote{The 11 sectors of the Global Industry Classification Standard are: Information Technology, Health Care, Financials, Consumer Discretionary, Communication Services, Industrials, Consumer Staples, Enery, Utilities, Real Estate, and Materials.}. 
    This is one possible data-driven proxy for the beta-ordered uniform partition. (Sector Separated)
    \item $\hat{h}^t_{L(\mathcal{P})}$, the MAPS estimator of $R^t$ where $\mathcal{P}$ is a beta ordered uniform partition with 11 atoms constructed by using the true ordering of the entries of $\beta_t$. (Beta Ordered)
\end{enumerate}

For each of these four choices of estimator $h^t$, we examine three different measures of error: the squared $\ell_2$ error $||h^t - b_t||^2$, the scaled squared tracking error $p\mathcal{T}^2(h^t)$, and the scaled optimization bias $p\mathcal{E}^2_p(h^t)$.

Since we are interested in expected outcomes, we repeat the above experiment 100 times, and take the average of the errors as a monte carlo estimate of the expectations
\[
\bE[||h^t - b_t||^2], \quad \bE[p\mathcal{T}^2(h^t)], \quad \bE[p\mathcal{E}^2_p(h^t)],
\]
once for each $t$.
We then display box plots for the resulting distribution of 24 expected errors of each type, corresponding to the 24 historical betas.

\begin{figure}[!htb]
    \centering
    \begin{subfigure}[b]{0.3\textwidth}
    \centering
    \includegraphics[width=\textwidth]{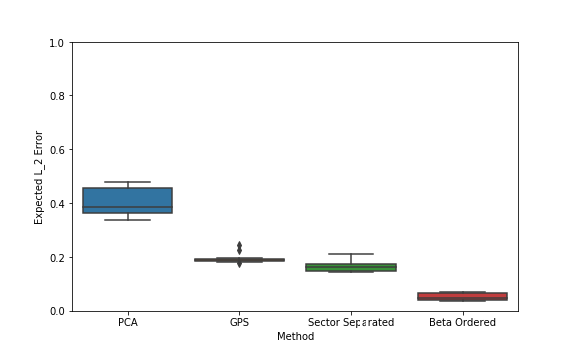}
    \caption{$\ell_2$ error}
    \label{fig:empricial-singleL2}
    \end{subfigure}
    \hfill
    \begin{subfigure}[b]{0.3\textwidth}
    \centering
    \includegraphics[width=\textwidth]{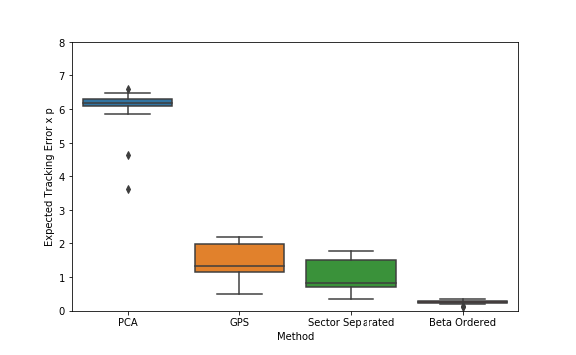}
    \caption{tracking error}
    \label{fig:empirical-singletracking}
    \end{subfigure}
    \hfill
    \begin{subfigure}[b]{0.3\textwidth}
    \centering
    \includegraphics[width=\textwidth]{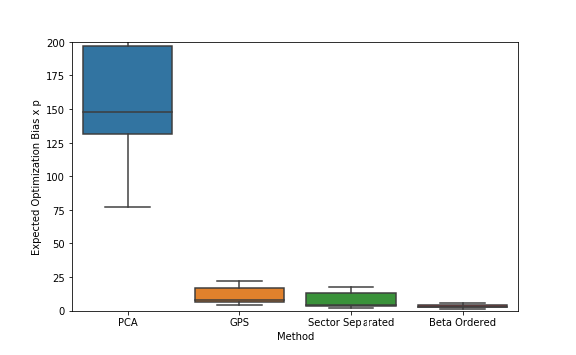}
    \caption{optimization bias}
    \label{fig:empirical-singleoptbias}
    \end{subfigure}
    \caption{Box plots summarizing the distribution of 24 monte carlo-estimated expected errors for the PCA, GPS, Sector Separated, and Beta Ordered estimators (left to right in each figure). The experiment is conducted over $488$ S$\&$P 500 companies. This experiment reveals that the Sector Separated estimator is able to capture some of the ordering information and therefore outperforms the GPS estimator. The Beta Ordered estimator performs best.}
    \label{fig:empiricalsingleblock}
\end{figure}

Figure \ref{fig:empiricalsingleblock} shows a similar story for all three error measures.  The GPS estimator significantly outperforms the PCA estimator, and the Beta Ordered estimator, which assumes the ability to rank order partition the betas, is significantly the best. 

The result of more interest is that a sector partition approximates a beta ordered partition well enough to improve on the GPS estimate. This approach takes advantage of the fact that betas of stocks in a common sector tend on average to be closer to each other than to betas in other sectors. The Sector Separated MAPS estimator does not require any information not easily available to the practitioner, and so represents a costless improvement on the GPS estimation method.

\subsubsection{Double Data Block}
In order to test the dynamical MAPS estimator that is designed to take advantage of serial correlation in the betas, we will generate a test bed of double data blocks of simulated market observations using the 24 WRDS historical betas for the same time period as before. 

For each $t = 1,2, \dots, 12$, we generate 12 simulated monthly market returns for $\beta_t$ and $\beta_{t+12}$ according to
\begin{equation}
     R_i^t=\beta_t X_i^t+ Z_{i}^t, \quad i=1,2,...,12
\end{equation}
\begin{equation}
     R_i^{t+12}=\beta_{t+12} X_i^{t+12}+ Z_{i}^{t+12}, \quad i=1,2,...,12
\end{equation}
were $X$ and $Z$ are generated independently as before.

This provides, for each $t$, two $p \times 12$ ``single block" returns matrices
$R^t$ and $R^{t+12}$ each covering 12 months, and a combined ``double block" $p \times 2n$ returns matrix $R_t = [R^t R^{t+12}]$ containing 24 consecutive monthly returns of the $p$ stocks.

The estimation problem, given observation of the double block of data $R_{t}$, is to estimate the normalized beta vector $\beta^{t+12}/||\beta^{t+12}||$ corresponding to the most recent 12 months.
This estimate then implies an estimated covariance
matrix for that 12 month period according to equations (\ref{eq:estimatedcovmatrix}) and (\ref{eq:estimatedscalars}), and allows us to measure the estimation error as before.

 
 We compare the following estimators:
\begin{itemize}
    \item $h_{PCA}^s$, the PCA estimator of $R^{t+12}$.
    \item $h_{PCA}^d$, the PCA estimator of the double block $R_t=[R^{t} R^{t+12}]$.
    \item $\hat{h}^s_q$ the GPS estimator of $R^{t+12}$.
    \item $\hat{h}^d_q$ the GPS estimator of $R_t=[R^{t} R^{t+12}]$.
    \item $\hat{h}_{L{_D}}$ the dynamical MAPS estimator defined on the double block $R_t=[R^{t} R^{t+12}]$ by Equation (\ref{eq:dynamicalmapsestimator}).
\end{itemize}

We will report our results using the same three error metrics as before: $\mathbb{E}[||*-b||^2]$, $\mathbb{E}[p \mathcal{T}^2(*)]$, $\mathbb{E}[p\mathcal{E}^2_p(*)]$ for each of the five estimators.  
To obtain estimated expectations, we repeat the experiments 100 times and compute the average.  The box plots summarize the distribution of the 12 overlapping double block expected errors.

The experiment shows that the Dynamical MAPS estimator outperforms the others, and illustrates the promise of Theorem \ref{T4}, which is based on the hypothesis that betas exhibit some serial correlation.  Another benefit of the Dynamical MAPS approach is to relieve  the practitioner from the burden of choosing whether to use a GPS1 or GPS2 estimator when a double block of data is available.

\begin{figure}[!htb]
    \centering
    \begin{subfigure}[b]{0.30\textwidth}
    \centering
    \includegraphics[width=\textwidth]{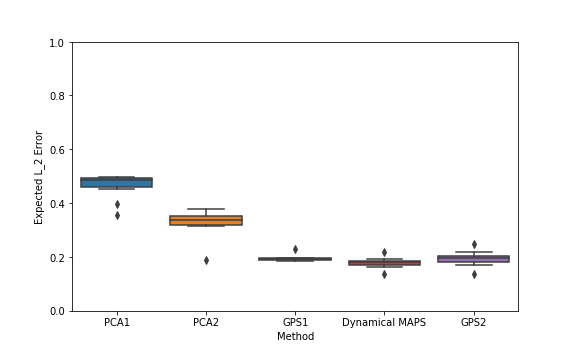}
    \caption{$\ell_2$ error}
    \label{fig:empricial-doubleL2}
    \end{subfigure}
    \hfill
    \begin{subfigure}[b]{0.30\textwidth}
    \centering
    \includegraphics[width=\textwidth]{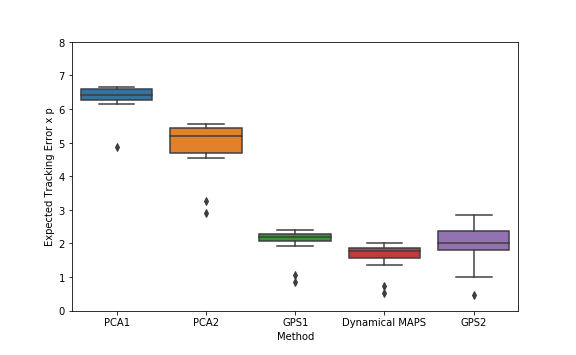}
    \caption{tracking error}
    \label{fig:empirical-doubletracking}
    \end{subfigure}
    \hfill
    \begin{subfigure}[b]{0.30\textwidth}
    \centering
    \includegraphics[width=\textwidth]{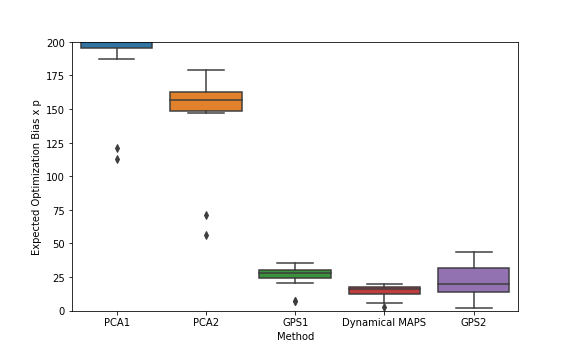}
    \caption{optimization bias}
    \label{fig:empirical-doubleoptbias}
    \end{subfigure}
    \caption{Box plots for three kinds of expected error for (left to right) the PCA1, PCA2, GPS1,  Dynamical MAPS, and GPS2 estimators, summarizing the distribution of 12 different expected errors for each estimator corresponding to 12 consecutive months of 2020.  The experiment is conducted using $488$ S$\&$P 500 companies. }
    \label{fig:empiricaldoubleblock}
\end{figure}

%% file: maps-mainproofs-extended.tex
The proofs of the main theorems proceed by means of some intermediate results involving an ``oracle estimator", defined in terms of the unobservable $b$ but equal to the MAPS estimator in the asymptotic limit (Theorem \ref{thm1} below).  Several technical supporting propositions and lemmas are needed and postponed to Section \ref{sec:supplement}.

\subsection{Oracle Theorems}

A key tool in the proofs is the {\em oracle estimator} $h_L$, which is a version of $\hat h_L$ but defined in terms of $b$, our estimation target.

Given a subspace $L = L_p$ of $\bR^p$, we define

\begin{equation}
    h_L=\frac{\underset{<h,L>}{\operatorname{proj}}(b)}{||\underset{<h,L>}{\operatorname{proj}}(b)||}.
    \label{oracle}
\end{equation}

Here $<h,L>$ denotes the span of $h$ and $L$, and note that if $L=\{0\}$ we get $h_L=h$, the PCA estimator. A nontrivial example for the selection would be $L_p=<q>$, which generates $h_q$, the oracle version of the GPS estimator in \cite{goldberg2018}.
The following theorem says that asymptotically the oracle estimator (\ref{oracle}) converges to the MAPS estimator (\ref{data}).

\begin{thm}\label{thm1}
Let the assumptions \textbf{1},\textbf{2},\textbf{3} and \textbf{4} hold. Suppose $\{L_p\}$ be any sequence of random linear subspaces that is independent of the entries of $Z$, such that $dim(L_p)$ is a square root dominated sequence. Then
\begin{equation}
    \lim\limits_{p\rightarrow \infty}||\hat{h}_L-h_L||=0 .
\end{equation}
\end{thm}

The proof of Theorem \ref{thm1} requires the following proposition, proved in Section \ref{sec:supplement}.

\begin{prop}\label{t5}
Under the assumptions of Theorem \ref{thm1}, let $h = h_{PCA}$ be the PCA estimator, equal to the unit leading eigenvector of the sample covariance matrix.
Then, almost surely:
\begin{enumerate}
    \item $\lim\limits_{p\rightarrow \infty}\big((h,\underset{L}{\operatorname{proj}}(h))-(h,b)^2(b,\underset{L}{\operatorname{proj}}(b))\big)=0$, \label{new1}
    \item $\lim\limits_{p\rightarrow \infty}\big((b,\underset{L}{\operatorname{proj}}(h))-(h,b)(b,\underset{L}{\operatorname{proj}}(b)))\big)=0$, \label{new2} \text{ and }
    \item $\lim\limits_{p\rightarrow \infty}||\underset{L}{\operatorname{proj}}(h)-(h,b)\underset{L}{\operatorname{proj}}(b)||=0.$ \label{new3}
\end{enumerate}
In particular, $\frac{\underset{L}{\operatorname{proj}}(h)}{||\underset{L}{\operatorname{proj}}(h)||}$ converges asymptotically to $\frac{\underset{L}{\operatorname{proj}}(b)}{||\underset{L}{\operatorname{proj}}(b)||}$.
\end{prop}

\begin{proof}[\text{Proof of the Theorem \ref{thm1}}:]

Recall from (\ref{data}) that,
\begin{equation}
    \hat{h}_L=\frac{\tau_ph+\underset{L}{\operatorname{proj}}(h)}{||\tau_ph+\underset{L}{\operatorname{proj}}(h)||} \text{ }\text{ where }\text{ } \tau_p=\frac{\psi_p^2-||\underset{L}{\operatorname{proj}}(h)||^2}{1-\psi_p^2} \nonumber.
\end{equation}
By Lemma \ref{lem:psi-infty}, $\psi_p$ has an almost sure limit $\psi_{\infty} = (h,b)_{\infty} \in (0,1)$, and hence $\tau_p$
is bounded in $p$ almost surely.

Let $\Omega_1\subset \Omega$ be the almost sure set for which the conclusions of Proposition \ref{t5} hold. Define the notation
\[
a_p(\omega)=||\hat{h}_{L_p}-h_{L_p}||
\]
and  
\begin{equation}
    \gamma_p=\frac{(h,b)-(b,\underset{L}{\operatorname{proj}}(h))}{1-||\underset{L}{\operatorname{proj}}(h)||^2} \nonumber.
\end{equation}

The proof will follow steps 1-4 below:
\begin{enumerate}
    \item For every $\omega\in \Omega_1$ and sub-sequence $\{p_k\}_{k=1}^{\infty}\subset \{p\}_{1}^{\infty}$ satisfying $$\limsup\limits_{k\rightarrow \infty}||\underset{L_{p_k}}{\operatorname{proj}}(b)||(\omega)<1$$ 
    we prove 
    \begin{equation}
 0< \liminf\limits_{k\rightarrow\infty}\gamma_{p_k}(\omega)\leq \limsup\limits_{k\rightarrow\infty}\gamma_{p_k}(\omega) < \infty \nonumber
\end{equation}
and
\begin{equation}
 0< \liminf\limits_{k\rightarrow\infty}\tau_{p_k}(\omega)\leq \limsup\limits_{k\rightarrow\infty}\tau_{p_k}(\omega) < \infty \nonumber.
\end{equation}
\item For every $\omega\in \Omega_1$ and sub-sequence $\{p_k\}_{k=1}^{\infty}\subset \{p\}_{1}^{\infty}$ satisfying $$\limsup\limits_{k\rightarrow \infty}||\underset{L_{p_k}}{\operatorname{proj}}(b)||(\omega)<1$$
we use step 1 to   prove   
$\lim\limits_{k\rightarrow\infty}a_{p_k}(w)$=0
\item Set $\Omega_0=\{\omega\in\Omega \big| \limsup\limits_{p\rightarrow\infty}||\underset{L_{p}}{\operatorname{proj}}(b)||^2=1\}$. 
Fix $\omega\in \Omega_0\cap\Omega_1$ and prove using step {2} that $\lim\limits_{p\rightarrow\infty}a_{p}(\omega)=0$
\item  Finish the proof by applying step 2 for all $\omega\in \Omega_0^c\cap \Omega_1$ when $\{p_k\}$ is set to $\{p\}$. 
\end{enumerate}

\textbf{ Step  1}: Since $\omega\in \Omega_1$ we have the following immediate implications of  Proposition \ref{t5},
\begin{equation}
  \limsup\limits_{k\rightarrow\infty}||\underset{L_{p_k}}{\operatorname{proj}}(h)||^2=(h,b)_{\infty}^2\limsup\limits_{k\rightarrow\infty}||\underset{L_{p_k}}{\operatorname{proj}}(b)||^2.  \label{eq:1} 
\end{equation}
\begin{equation}
    \limsup\limits_{k\rightarrow\infty} (b,\underset{L_{p_k}}{\operatorname{proj}}(h))=(h,b)_{\infty}\limsup\limits_{k\rightarrow\infty}||\underset{L_{p_k}}{\operatorname{proj}}(b)||^2. \label{eq:2}
\end{equation}
Using the assumption $\limsup\limits_{k\rightarrow\infty}||\underset{L_{p_k}}{\operatorname{proj}}(b)||^2<1$, we update (\ref{eq:1}) and (\ref{eq:2}) as,
\begin{equation}
    \limsup\limits_{k\rightarrow\infty}||\underset{L_{p_k}}{\operatorname{proj}}(h)||^2< (h,b)_{\infty}^2<1 \label{eq:3}
\end{equation}
\begin{equation}
    \limsup\limits_{k\rightarrow\infty} (b,\underset{L_{p_k}}{\operatorname{proj}}(h))<(h,b)_{\infty} \label{eq:4}
\end{equation}
for the given $\omega\in \Omega_1$. We can use (\ref{eq:3}) on the numerator of $\tau_{p_k}$ to show,
\begin{align*}
    \liminf\limits_{k\rightarrow\infty}\big(\psi_{p_k}^2-||\underset{L_{p_k}}{\operatorname{proj}}(h)||\big)
    &\geq \liminf\limits_{k\rightarrow\infty}\psi_{p_k}^2-\limsup\limits_{k\rightarrow\infty}||\underset{L_{p_k}}{\operatorname{proj}}(h)||^2 \\
    &=(h,b)^2_{\infty}-\limsup\limits_{k\rightarrow\infty}||\underset{L_{p_k}}{\operatorname{proj}}(h)||^2>0.
\end{align*}
That together with the fact that the denominator of $\tau_{p_k}$ has a limit in $(0,\infty)$ implies, 

\begin{equation}
    0< \liminf\limits_{k\rightarrow\infty}\tau_{p_k}(\omega)\leq \limsup\limits_{k\rightarrow\infty}\tau_{p_k}(\omega) < \infty \label{tau}
\end{equation}
Similarly we can use {(\ref{eq:4})} on the numerator of $\gamma_{p_k}$ as,
\begin{equation}
    \liminf\limits_{k\rightarrow\infty}\big((h,b)-(b,\underset{L_{p_k}}{\operatorname{proj}}(h))\big) 
    \geq (h,b)_{\infty}-\limsup\limits_{k\rightarrow\infty}(b,\underset{L_{p_k}}{\operatorname{proj}}(h)) >0. \label{eq:5}
\end{equation}
Also {(\ref{eq:3})} can be used on the denominator of $\gamma_{p_k}$ as,
\begin{equation}
    \liminf\limits_{k\rightarrow \infty}{1-||\underset{L_{p_k}}{\operatorname{proj}}(h)||^2}>1-\limsup\limits_{k\rightarrow \infty}||\underset{L_{p_k}}{\operatorname{proj}}(h)||^2>0 \label{eq:6}
\end{equation}
Using {(\ref{eq:5})} and {(\ref{eq:6})} we get,
\begin{equation}
 0< \liminf\limits_{k\rightarrow\infty}\gamma_{p_k}(\omega)\leq \limsup\limits_{k\rightarrow\infty}\gamma_{p_k}(\omega) < \infty \label{gamma}
\end{equation}
for the given $\omega\in\Omega_1$. This completes the {step  1}.
\\\\
\textbf{Step  2:} We have the following initial observation,
\begin{equation}
    1\geq ||\underset{<h,L_{p_k}>}{\operatorname{proj}}(b)||\geq ||\underset{<h>}{\operatorname{proj}}(b)||=(h,b)
\end{equation}
and using that we get $$1\geq \limsup\limits_{p\rightarrow}||\underset{<h,L_{p_k}>}{\operatorname{proj}}(b)||\geq \liminf\limits_{p\rightarrow}||\underset{<h,L_{p_k}>}{\operatorname{proj}}(b)|| \geq (h,b)_{\infty}>0.$$
Given that, in order to show $\lim\limits_{k\rightarrow\infty}a_{p_k}(\omega)=0$, it suffices to show $\tau_{p_k}h+\underset{L_{p_k}}{\operatorname{proj}}(h)$ converges to a scalar multiple of  $\underset{<h,L_{p_{k}}>}{\operatorname{proj}}(b)$ since that scalar clears after normalizing the vectors. To motivate that lets re-write $\underset{<h,L_{p_k}>}{\operatorname{proj}}(b)$ as,
\begin{align}
    \underset{<h,L_{p_k}>}{\operatorname{proj}}(b)
    &=\underset{<h-\underset{L_{p_k}}{\operatorname{proj}}(h),L_{p_k}>}{\operatorname{proj}}(b) \nonumber\\
    &=\underset{L_{p_k}}{\operatorname{proj}}(b)+\bigg(\frac{h-\underset{L_{p_k}}{\operatorname{proj}}(h)}{||h-\underset{L_{p_k}}{\operatorname{proj}}(h)||},b\bigg)\frac{h-\underset{L_{p_k}}{\operatorname{proj}}(h)}{||h-\underset{L_{p_k}}{\operatorname{proj}}(h)||} \nonumber \\
    &=\underset{L_{p_k}}{\operatorname{proj}}(b)+\gamma_{p_k} (h-\underset{L_{p_k}}{\operatorname{proj}}(h)) \label{7.5} \\
    &=\gamma_{p_k}(h+\frac{1}{\gamma_{p_k}}\underset{L_{p_k}}{\operatorname{proj}}(b)-\underset{L_{p_k}}{\operatorname{proj}}(h)). \label{8}
\end{align}
We also have,
\begin{equation}
    \tau_{p_k}h+\underset{L_{p_k}}{\operatorname{proj}}(h)=\tau_{p_k}(h+\frac{1}{\tau_{p_k}}\underset{L_{p_k}}{\operatorname{proj}}(h)). \label{9}
\end{equation}
Since we have $\tau_{p_k}$ and $\gamma_{p_k}$ satisfying {(\ref{tau})} and {(\ref{gamma})} respectively, we have the equations {(\ref{8})} and {(\ref{9})} well defined asymptotically, which is sufficient for our purpose. Hence, from the above argument it is sufficient to show the convergence of $h+\frac{1}{\tau_{p_k}}\underset{L_{p_k}}{\operatorname{proj}}(h)$ to $h+\frac{1}{\gamma_{p_k}}\underset{L_{p_k}}{\operatorname{proj}}(b)-\underset{L_{p_k}}{\operatorname{proj}}(h)$. That is equivalent to showing $\frac{1}{\tau_{p_k}}\underset{L_{p_k}}{\operatorname{proj}}(h)$ converges to $\frac{1}{\gamma_{p_k}}\underset{L_{p_k}}{\operatorname{proj}}(b)-\underset{L_{p_k}}{\operatorname{proj}}(h)$. We can re-write the associated quantity as,
\begin{equation}
    \big|\frac{1}{\tau_{p_k}}\underset{L_{p_k}}{\operatorname{proj}}(h)-\big(\frac{1}{\gamma_{p_k}}\underset{L_{p_k}}{\operatorname{proj}}(b)-\underset{L_{p_k}}{\operatorname{proj}}(h)\big)\big|=\big|(1+\frac{1}{\tau_{p_k}})\underset{L_{p_k}}{\operatorname{proj}}(h)-\frac{1}{\gamma_{p_k}}\underset{L_{p_k}}{\operatorname{proj}}(b)\big| \label{10}
\end{equation}
Using Proposition \ref{t5} part 3 in {(\ref{10})}, it is equivalent to prove \\
$\big|(1+\frac{1}{\tau_{p_k}})(h,b)-\frac{1}{\gamma_{p_k}}\big|$ converges to $0$.  We re-write it as
\begin{align}
    |(\frac{1}{\tau_{p_k}}+1)(h,b)-\frac{1}{\gamma_{p_k}}|
    &=\bigg|\frac{(h,b)(1-||\underset{L_{p_k}}{\operatorname{proj}}(h)||^2}{\psi_{p_k}^2-||\underset{L_{p_k}}{\operatorname{proj}}(h)||^2}-\frac{1-||\underset{L_{p_k}}{\operatorname{proj}}(h)||^2}{(h,b)-(\underset{L_{p_k}}{\operatorname{proj}}(h),b)}\bigg| \nonumber\\
    &=|1-||\underset{L_{p_k}}{\operatorname{proj}}(h)||^2|\bigg|\frac{(h,b)}{\psi_{p_k}^2-||\underset{L_{p_k}}{\operatorname{proj}}(h)||^2}-\frac{1}{(h,b)-(\underset{L_{p_k}}{\operatorname{proj}}(h),b)}\bigg|
    \label{final}
\end{align}
Using parts ({1}) and ({2}) of Proposition \ref{t5} and the fact that $\psi_{p_k}^2$ converges to $(h,b)_{\infty}^2$ shows that ({\ref{final}}) converges to $0$ for the given $\omega\in \Omega_1$. This completes {step 2}.
\medskip

\noindent
\textbf{Step 3:} Fix $\omega\in \Omega_0\cap\Omega_1$. 
To show that $\lim_{p \to \infty} a_p(\omega) = 0$, it suffices to show that for any sub-sequence $\{p_k\}_{k=1}^{\infty}\subset \{p\}_{1}^{\infty}$ there exist a further sub-sequence $\{s_t\}_{t=1}^{\infty}$ such that $\lim\limits_{t\rightarrow \infty}a_{s_t}(\omega)=0$. Let  $\{p_k\}_{k=1}^{\infty}$ be a subsequence. We have one of the following cases,
\begin{equation}
    \limsup\limits_{k\rightarrow\infty}||\underset{L_{p_k}}{\operatorname{proj}}(b)||(\omega)^2<1 \nonumber
\end{equation} 
or
\begin{equation}
 \limsup\limits_{k\rightarrow\infty}||\underset{L_{p_k}}{\operatorname{proj}}(b)||(\omega)^2=1   \nonumber
\end{equation}

If it is strictly less than $1$, then we get from the {step 2} that $\lim\limits_{k\rightarrow\infty}a_{p_k}(\omega)=0$. In that case we take the further sub-sequence of  equal to $\{p_k\}$. 

If it is equal to $1$, then we get a further sub-sequence $\{s_t\}$ s.t \\ $\lim\limits_{t\rightarrow\infty}||\underset{L_{s_t}}{\operatorname{proj}}(b)||^2=1$. Using this and Proposition \ref{t5} we get the following,
\begin{equation}
    \lim\limits_{t\rightarrow\infty} ||\underset{L_{s_t}}{\operatorname{proj}}(h)||^2=(h,b)^2_{\infty} \text{ }\text{ and }  \lim\limits_{t\rightarrow\infty}(b,\underset{L_{s_t}}{\operatorname{proj}}(h))=(h,b)_{\infty} \nonumber
\end{equation}
which implies $ \lim\limits_{t\rightarrow\infty}\tau_{s_t}(\omega)= \lim\limits_{t\rightarrow\infty}\gamma_{s_t}(\omega)=0$. Using this on the definition of $\hat{h}_L$ and the equation {(\ref{7.5})} we get,
\begin{equation}
     \lim\limits_{t\rightarrow\infty}\big|\big|\hat{h}_{L_{s_t}}-\frac{\underset{L_{s_t}}{\operatorname{proj}}(h)}{||\underset{L_{s_t}}{\operatorname{proj}}(h)||}\big|\big|=0 \text{ }\text{ and }\text{ } \lim\limits_{t\rightarrow\infty}\big|\big|h_{L_{s_t}}-\frac{\underset{L_{s_t}}{\operatorname{proj}}(b)}{||\underset{L_{s_t}}{\operatorname{proj}}(b)||}\big|\big|=0 \label{11}
\end{equation}
We can now decompose $a_{s_t}=||\hat{h}_{L_{s_t}}-h_{L_{s_t}}||$ into familiar components via the triangle inequality as follows,
\begin{align*}
    a_{s_t}=||\hat{h}_{L_{s_t}}-h_{L_{s_t}}||\leq \big|&\big|\hat{h}_{L_{s_t}}-\frac{\underset{L_{s_t}}{\operatorname{proj}}(h)}{||\underset{L_{s_t}}{\operatorname{proj}}(h)||}\big|\big|+\big|\big|h_{L_{s_t}}-\frac{\underset{L_{s_t}}{\operatorname{proj}}(b)}{||\underset{L_{s_t}}{\operatorname{proj}}(b)||}\big|\big|
    \nonumber\\
    &+\big|\big|\frac{\underset{L_{s_t}}{\operatorname{proj}}(b)}{||\underset{L_{s_t}}{\operatorname{proj}}(b)||}- \frac{\underset{L_{s_t}}{\operatorname{proj}}(h)}{||\underset{L_{s_t}}{\operatorname{proj}}(h)||}\big|\big| 
\end{align*}
Using {(\ref{11})}, we know that the first and the second terms on the right hand side converge to $0$ for the given $\omega\in \Omega_0\cap\Omega_1$. Since we have $\lim\limits_{t\rightarrow\infty} ||\underset{L_{s_t}}{\operatorname{proj}}(h)||^2=(h,b)^2_{\infty}$ and $\lim\limits_{t\rightarrow\infty} ||\underset{L_{s_t}}{\operatorname{proj}}(b)||^2=1$, proving the third term on the right hand side converges to $0$ is equivalent to proving
\begin{equation}
   \lim\limits_{t\rightarrow\infty} \big|\big|  \underset{L_{s_t}}{\operatorname{proj}}(h)-(h,b) \underset{L_{s_t}}{\operatorname{proj}}(b)\big|\big|=0, \nonumber
\end{equation}
which is true by Proposition \ref{t5}. This completes the {step 3}. 
\\\\
\textbf{Step 4:} In {step 3} we proved the theorem for every $\omega\in \Omega_0\cap\Omega_1$. Replacing $\{p_k\}$ in {step 2} by the whole sequence of indices $\{p\}$, we get the theorem for every $\omega\in \Omega_0^c\cap\Omega_1$. These together shows that we have,
\begin{equation}
    \lim\limits_{p\rightarrow\infty} a_p(w)=0 \text{ }\text{ for all }\omega\in \Omega_1  
    \nonumber
\end{equation}
which completes the proof of Theorem \ref{thm1}.
\end{proof}

\subsection{Proof of Theorem \ref{T1}}

The proof of Theorem \ref{T1}(a) is an immediate application of Theorem \ref{thm1}.

\begin{proof}[\text{Proof of the Theorem \ref{T1}(a)}:]

From the definitions of $h_L$ and $h_q$, and as long as $q \in L_p$, we have
\begin{equation}
    ||h_{L_p}-b||\leq ||h_q-b|| \nonumber
\end{equation} 
and therefore
\begin{eqnarray*}
||\hat h_{L_p} - b|| &\leq& ||\hat h_{L_p} - h_{L_p}|| + ||h_{L_p} - b||\\
 &\leq&  ||\hat h_{L_p} - h_{L_p}|| + ||h_{q} - b|| \\
 &\leq& ||\hat h_{L_p} - h_{L_p}|| + ||\hat h_{q} - b|| 
\end{eqnarray*}
since $||h_{q} -b|| \leq ||\hat h_{q} -b||$ for all $p$.
Applying Theorem \ref{thm1} gives
\begin{equation}
\limsup ||\hat h_{L_p} - b|| \leq ||\hat h_{q} -b||_{\infty}. \nonumber
\end{equation}

\end{proof}

To prove the remainder of Theorem  \ref{T1} we need the following intermediate result concerning Haar random subspaces, proved in Section \ref{sec:supplement}.

\begin{prop}\label{t7}
Suppose, for each $p$, $z_p$ is a (possibly random) point in $\mathbb{S}^{p-1}$ and  
$\mathcal{H}_p$ is a Haar random subspace of $\bR^p$ that is independent of $z_p$. Assume the
sequence $\{ \dim \mathcal{H}_p \}$ is square root dominated.

 Then 
    \begin{equation}
        \lim\limits_{p\rightarrow \infty}||\underset{\mathcal{H}_p}{\operatorname{proj}}(z_p)||^2=0 \text{ almost surely.} \nonumber
    \end{equation} 
\end{prop}

\begin{proof}[\text{Proof of the Theorem \ref{T1}(b,c)}:]
Theorem \ref{thm1} is applicable. Hence, it suffices to prove the results for the oracle version of the MAPS estimator. 

Since the scalars clear after normalization, it suffices to prove the following assertions,
\begin{equation}
    \lim\limits_{p\rightarrow\infty} ||\underset{<h,\mathcal{H}>}{\operatorname{proj}}(b)-\underset{<h>}{\operatorname{proj}}(b)||_2=0 \label{d4}
\end{equation}
and
\begin{equation}
    \lim\limits_{p\rightarrow\infty} ||\underset{<h,q,\mathcal{H}>}{\operatorname{proj}}(b)-\underset{<h,q>}{\operatorname{proj}}(b)||_2=0. \label{d5}
\end{equation}
We first consider ({\ref{d4}}), rewriting the left hand side as
\begin{align}
&\lim\limits_{p\rightarrow \infty} ||\underset{\mathcal{H}}{\operatorname{proj}}(b)+\underset{h-\underset{\mathcal{H}}{\operatorname{proj}}(h)}{\operatorname{proj}}(b)-\underset{<h>}{\operatorname{proj}}(b)||_2 
\nonumber\\
&\leq ||\underset{\mathcal{H}}{\operatorname{proj}}(b)||_2+||\underset{h-\underset{\mathcal{H}}{\operatorname{proj}}(h)}{\operatorname{proj}}(b)-\underset{<h>}{\operatorname{proj}}(b)||_2 \label{d6}
\end{align}
The first term of (\ref{d6}) converges to $0$ by setting $z=b$ in Proposition \ref{t7}. Moreover,  Propositions \ref{t7} and \ref{t5} imply $\underset{\mathcal{H}}{\operatorname{proj}}(h)$ converges to the origin in the $l_2$ norm. Hence we have $h-\underset{\mathcal{H}}{\operatorname{proj}}(h)$ is converging to $h$ in $l_2$ norm. That implies the second term in ({\ref{d6}}) converges to 0, which in turn proves ({\ref{d4}}). 

Next, rewrite the expression in the assertion ({\ref{d5})} as,
\begin{align}
    &||\underset{\mathcal{H}}{\operatorname{proj}}(b)+\underset{<h-\underset{\mathcal{H}}{\operatorname{proj}}(h),q-\underset{\mathcal{H}}{\operatorname{proj}}(q)>}{\operatorname{proj}}(b)-\underset{<h,q>}{\operatorname{proj}}(b)||\nonumber \\
    &\leq ||\underset{\mathcal{H}}{\operatorname{proj}}(b)||+ ||\underset{<h-\underset{\mathcal{H}}{\operatorname{proj}}(h),q-\underset{\mathcal{H}}{\operatorname{proj}}(q)>}{\operatorname{proj}}(b)-\underset{<h,q>}{\operatorname{proj}}(b)||\label{d7}
\end{align}
Similarly the first term of  (\ref{d7}) converges to $0$  by Proposition \ref{t7}. Note that \ref{t7} also applies when we set $z=q$, and hence  $\underset{\mathcal{H}}{\operatorname{proj}}(q)$ converges to the origin in the $l_2$ norm. Hence the basis elements of
$<h-\underset{\mathcal{H}}{\operatorname{proj}}(h),q-\underset{\mathcal{H}}{\operatorname{proj}}(q)>$ converge to the basis elements of $<h,q>$, which implies the second term of ({\ref{d7}}) converges to $0$ as well. That completes the proof.
\end{proof}

\subsection{Proof of Theorem \ref{T3}}

We need the following lemma.
\begin{lem}\label{l5}
Let $\mathcal{P}(p)$ be a sequence of uniform $\beta$-ordered partitions such that $\lim\limits_{p\rightarrow\infty}|\mathcal{P}(p)|=\infty$.  Then for $L_p=L(\mathcal{P}(p))$ we have,
\begin{equation}
    \lim\limits_{p\rightarrow \infty} ||\underset{L}{\operatorname{proj}}(b)||=1
\end{equation}
almost surely.
\end{lem}
\begin{proof}
To be more precise about $L=L(\mathcal{P})$, set $\mathcal{P}(p)=\{I_1,I_2,...,I_{k_p}\}$ and denote the defining basis of the corresponding subspace $L_p=L(\mathcal{P})$ by  the orthonormal set $\{v_1,v_2,...,v_{k_p}\}$.  

Then

\begin{align}
    1-||\underset{L}{\operatorname{proj}}(b)||^2
    &=1-\lim\limits_{p\rightarrow \infty}\sum\limits_{i=1}^{k_p}(b,v_i)^2 \nonumber \\
    &=\sum\limits_{i=1}^{p}b_i^2-\lim\limits_{p\rightarrow \infty}\sum\limits_{i=1}^{k_p}(b,v_i)^2 \nonumber \\
    &=\lim\limits_{p\rightarrow \infty}\frac{1}{||\beta||^2}\sum\limits_{i=1}^{k_p}(\sum\limits_{j\in I_i}\beta_j^2-\frac{1}{|I_i|}(\sum\limits_{j\in I_i}\beta_i)^2)  \nonumber\\
    &=\lim\limits_{p\rightarrow \infty}\frac{1}{||\beta||^2}\sum\limits_{i=1}^{k_p}(\sum\limits_{j\in I_i}(\beta_j-\frac{1}{|I_i|}(\sum\limits_{j\in I_i}\beta_i))^2 
\end{align}

Now define the random variables $a_i=\underset{j\in I_i}{max}(\beta_j)$, $c_i=\underset{j\in I_i}{min}(\beta_j)$ for all $1\leq i \leq k_p$. Without loss of generality, $c_{k_p}\leq a_{k_p} \leq ... \leq c_1 \leq a_1$.    Since the sequence $\{\mathcal{P}(p)\}$ is uniform, there exists $M>0$ such that
\begin{equation}
      \max\limits_{I\in \mathcal{P}(p)}|I|\leq  \frac{Mp}{|\mathcal{P}(p)|}  .
    \end{equation}
Then
\begin{align}
    \lim\limits_{p\rightarrow \infty}\frac{1}{||\beta||^2}\sum\limits_{i=1}^{k_p}(\sum\limits_{j\in I_i}(\beta_j-\frac{1}{|I_i|}(\sum\limits_{j\in I_i}\beta_i))^2
    &\leq \lim\limits_{p\rightarrow \infty}\frac{1}{||\beta||^2}\sum\limits_{i=1}^{k_p}|I_i|(a_i-c_i)^2 \nonumber\\
    &\leq \lim\limits_{p\rightarrow \infty}\frac{\frac{Mp}{k_p}}{||\beta||^2}\sum\limits_{i=1}^{k_p}(a_i-c_i)^2 \\
    &= \lim\limits_{p\rightarrow \infty}\frac{M}{\frac{||\beta||^2}{p}}\frac{1}{k_p} (a_1-c_{k_p})^2 
    \label{e27}
\end{align}
The term $a_1-c_{k_p}$ appearing in (\ref{e27}) is uniformly bounded since the $\beta$'s are uniformly bounded. Also, $\frac{||\beta||^2}{p}$ is finite and away from zero asymptotically. Using those together with the fact that $\lim\limits_{p\rightarrow \infty}k_p=\infty$  we get the limit in (\ref{e27}) equal to $0$ for any realization of the random variables $\beta$. Note that this is stronger than almost sure convergence.
\end{proof}

\begin{proof}[\text{Proof of the Theorem \ref{T3}}:]
By an application of Theorem {\ref{thm1}} it suffices to prove the theorem for the oracle version of the MAPS estimator. Now
\begin{equation}
    ||b-\underset{<h,L>}{\operatorname{proj}}(b)||^2\leq ||b-\underset{L}{\operatorname{proj}}(b)||^2=1-||\underset{L}{\operatorname{proj}}(b)||^2 \label{d1}
\end{equation}
and note that application of Lemma \ref{l5} shows that $||\underset{L}{\operatorname{proj}}(b)||$ converges to 1 as $p$ tends to $\infty$. 
\end{proof}

\subsection{Proof of Theorem \ref{T4}(a)}

The proof of Theorem \ref{T4} requires the following proposition, from which part (a) of the theorem easily follows.  The proof of the proposition, along with the more difficult proof of the the strict inequality of \ref{T4} (b), appear in Section \ref{sec:supplement}.  

Recall that $h_1, h_2$ and $h$ are the PCA leading eigenvectors of the sample covariance matrices of the returns $R_1, R_2$ and $R$, respectively.

\begin{prop}\label{p1}
For each $p$ there is a vector $\Tilde{h}$ in the linear subspace $L \subset R^p$ generated by $h_1$ and $h_2$ such that $\lim\limits_{p\rightarrow\infty}||\Tilde{h}-h||=0$ almost surely.
\end{prop}

\begin{proof}[Proof of Theorem \ref{T4}(a)]

Since $dim(L_p)=2$ and $L_p=span(h_1,q)$ is independent of the asset specific portion $Z_2$ of the current block, Theorem {2.1} implies that $\hat{h}_L$ converges to $h_L$ almost surely in $l_2$ norm. Hence it suffices to establish the result for the oracle versions of the MAPS and the GPS estimators. 

Note
\begin{equation}
    (h_L,b)=||\underset{span(q,h_1,h_2)}{\operatorname{proj}}(b)||
\end{equation}
\begin{equation}
    (h_q^s,b)=||\underset{span(q,h_2)}{\operatorname{proj}}(b)||
\end{equation}
\begin{equation}
    (h_q^d,b)=||\underset{span(q,h)}{\operatorname{proj}}(b)||
\end{equation}
Using Proposition {\ref{p1}} we know there exist $\Tilde{h}\in  span(h_1,h_2)$ such that $\Tilde{h}$ converges to $h$ in $l_2$ almost surely. Since $span(q,\Tilde{h})\subset span(q,h_1,h_2)$, $$||\underset{span(q,h_1,h_2)}{\operatorname{proj}}(b)||\geq||\underset{span(q,\Tilde{h})}{\operatorname{proj}}(b)||.$$ Taking the limits of both sides we get
\begin{equation}
    \lim\limits_{p\rightarrow\infty} (h_L,b)=\lim\limits_{p\rightarrow\infty}||\underset{span(q,h_1,h_2)}{\operatorname{proj}}(b)||\geq \lim\limits_{p\rightarrow\infty}||\underset{span(q,h)}{\operatorname{proj}}(b)||=\lim\limits_{p\rightarrow\infty}(h^d_q,b). \label{a11}
\end{equation}
Similarly, since $span(q,h_1)\subset span(q,h_1,h_2)$,
\begin{equation}
    \lim\limits_{p\rightarrow\infty} (h_L,b)=\lim\limits_{p\rightarrow\infty}||\underset{span(q,h_1,h_2)}{\operatorname{proj}}(b)||\geq \lim\limits_{p\rightarrow\infty}||\underset{span(q,h_1)}{\operatorname{proj}}(b)||=\lim\limits_{p\rightarrow \infty}(h^d_q,b). \label{a12} 
\end{equation}
Inequalities ({\ref{a11}}) and ({\ref{a12}}) complete the proof of Theorem \ref{T4}(a).
\end{proof}

%% file: tech-proofs.tex
This section is devoted to four tasks: proving Proposition \ref{t5}, Proposition \ref{t7},  Proposition {\ref{p1}}, and part (b) of Theorem {\ref{T4}}. 

\subsection{Proof of Proposition \ref{t5}}

The following is equation ({33}) from \cite{goldberg2018}:
\begin{equation}
    h=\frac{\beta X^T\phi}{s_p\sqrt{n}}+\frac{Z \phi}{s_p\sqrt{n}} \label{core}
\end{equation}
where $s_p^2$ is the eigenvalue corresponding to $h$ and $\phi$ is the right singular vector of $\frac{1}{\sqrt{n}}R$. 
 For notational convenience we set 
 $$\Gamma_p=\frac{|\beta\| X^T\phi}{s_p\sqrt{n}}.$$
  
\begin{lem}[\cite{goldberg2018}]\label{l3}
Under the assumptions \textbf{1,2,3} and \textbf{4} we have 
\begin{enumerate}
\item $\lim\limits_{p\rightarrow \infty}\Gamma_p=(h,b)_{\infty}$,
\item $\lim\limits_{p\rightarrow\infty} \frac{s_p^2}{p}=\sigma_{\beta}^2+\frac{\delta^2}{n}$, and 
\item $\lim\limits_{p\rightarrow \infty}\phi_p=\frac{X}{|X|}$
\end{enumerate} 
almost surely.
\end{lem}


We need one more lemma to prepare for the proof of  Proposition \ref{t5}. 
Let $Z_k\in \mathbb{R}^p$ denote the $k$th column of $Z$. From our assumptions we know 
\begin{itemize}
    \item for each $k = 1, \dots, n$, $\{Z_{k}(i)=Z_{ik}:i\in \{1,2,...,p\}\}$  is an independent set of mean zero random variables, and
    \item there exists $M\in (0,\infty)$ such that $\mathbb{E}[Z_{k}(i)^{4}]\leq M$ for every $i\in\{1,2,...,p\}$.
\end{itemize}
\begin{lem}\label{l6}
Let $k \in \{1, \dots, n\}$ and let $u$ be a random unit vector of dimension $p$ and independent of $Z_{k}$. Then
\begin{equation}
    \mathbb{E}[(u,Z_{k})]=0 \text{ and } \mathbb{E}[(u,Z_k)^{4}] \leq 3M.
\end{equation}
\end{lem}
\begin{proof}
For notational convenience  set $X=Z_{k}$. 

Let $\theta=(\theta_1,\theta_2, ..., \theta_p)\in \mathbb{N}^p$ where $\sum\limits_{i=1}^{p}\theta_i=4$. For that, define $u^{\theta}=\prod\limits_{i=1}^{p}u_i^{\theta_i}$. Define $X^{\theta}$ similarly. We have,
\begin{equation}
    \mathbb{E}[(u,X)^{4}]=\sum\limits_{\theta}\mathbb{E}[u^{\theta}X^{\theta}]. \label{d10}
\end{equation}
Since $X_i$'s are independent and have mean $0$ and $u_i$ are independent of $X_i$'s,  if $\theta$ vector contains an entry equals to $1$, we would get the corresponding term $\mathbb{E}[u^{\theta}X^{\theta}]$ vanish. Hence we continue from the sum in (\ref{d10}) as,
\begin{align}
    &=\frac{4!}{2!2!}\sum\limits_{i< j}\mathbb{E}[u_i^2u_j^2X_i^2X_j^2]+\sum\limits_{i}\mathbb{E}[u_i^4X_i^4] \nonumber \\
    &=6\sum\limits_{i< j}\mathbb{E}[u_i^2u_j^2]\mathbb{E}[X_i^2X_j^2]+\sum\limits_{i}\mathbb{E}[u_i^4]\mathbb{E}[X_i^4] \nonumber\\
    &\leq M\big(6\sum\limits_{i< j}\mathbb{E}[u_i^2u_j^2]+\sum\limits_{i}\mathbb{E}[u_i^4]\big) \label{d11}
\end{align}
The last inequality in (\ref{d11}) follows by the assumption $\mathbb{E}[X_i^4]<M$ and an application of the Cauchy-Shawardz inequality
\begin{equation*}
    \mathbb{E}[X_i^2X_j^2]\leq \sqrt{ \mathbb{E}[X_i^4]\mathbb{E}[X_j^4]}\leq M.
\end{equation*}
We can continue from (\ref{d11}) withd
\begin{equation}
    = 3M\mathbb{E}\big[(\sum\limits_{i=1}^{p} u_i^2)^2\big]-2M\mathbb{E}[\sum\limits_{i=1}^{p}u_i^4]\leq 3M\mathbb{E}\big[(\sum\limits_{i=1}^{p} u_i^2)^2\big]=3M
\end{equation}
completing the proof of Lemma \ref{l6}.
\end{proof}

\begin{lem}\label{l22}
For each $p$, let $L_p$ be a (possibly random) linear subspace of $\bR^p$, with dimension $k_p$, that is
independent of $Z^p$. Assume the sequence $\{k_p\}$ is square root dominated.
Let $Z^p_s$ denote the $s$th column of $Z^p$.

 Then, for any $s\in \{1,2,...,n\}$, 
\begin{equation}
    \lim\limits_{p\rightarrow\infty}\frac{1}{p}||\underset{L}{\operatorname{proj}}(Z^p_s)||^2=0\label{new8}
\end{equation}
almost surely.
\end{lem}

\begin{proof}
Note first that, for any $p$ where $k_p=0$ we have $||\underset{L}{\operatorname{proj}}(Z^p_s)||=0$ so without loss of generality we can assume $k_p>0$ for all $p$. Under that assumption, there exist an orthonormal basis $\{u_1,u_2,...,u_{k_p}\}$ of $L_p$ that is independent of $Z^p$. Then we can rewrite the expression in (\ref{new8}) as
\begin{align}
    \frac{1}{p}||\underset{L}{\operatorname{proj}}(Z^p_s)||^2 &=\frac{1}{p}\sum\limits_{i=1}^{k_p}(u_i,Z^p_s)^2 \label{new9}
\end{align}
 
Set $Y_i=(u_i,Z_s)$ and observe that $Y_i$ depends on the selection of the orthonormal basis but $\frac{1}{p}\sum\limits_{i=1}^{k_p}Y_i^2$ does not. Now see that the lemma \ref{l6} implies 
\begin{equation}
    \mathbb{E}[Y_i^4]<3M \label{eq: lemma6.5_1}
\end{equation} 
for each $i$. Using that together with an application of the Cauchy-Shawardz inequality shows,
\begin{equation}
\mathbb{E}[Y_i^2Y_j^2]\leq \sqrt{\mathbb{E}[Y_i^4]\mathbb{E}[Y_j^4]} < 3M \label{eq: lemma6.5_2}
\end{equation}
for any $i\neq j$. Reading \ref{eq: lemma6.5_2} together with \ref{eq: lemma6.5_1}, we see that the \ref{eq: lemma6.5_2} is true for any $i$ and $j$ combination including the case $i=j$. We want to prove,

\begin{equation}
    \lim\limits_{p\rightarrow \infty}\frac{1}{p}\sum\limits_{i=1}^{k_p}Y_i^2=0 \label{new 10}.
\end{equation}

Using an application of the Chebyshev's inequality, \ref{eq: lemma6.5_1} and \ref{eq: lemma6.5_2} we argue as follows,

\begin{align}
\mathbb{P}(||\underset{L}{\operatorname{proj}}(Z^p_s)||^2>\epsilon p)
&=\mathbb{P}(\sum\limits_{i=1}^{k_p} Y_i^2>\epsilon p) \nonumber\\
&<\frac{\mathbb{E}[(\sum\limits_{i=1}^{k_p}Y_i^2)^2]}{\epsilon^2p^2} \nonumber\\
&<\frac{3M k_p^2}{\epsilon^2p^2}\label{new14}
\end{align}
\\
Note that the event $A_p:=\{||\underset{L}{\operatorname{proj}}(Z^p_s)||^2>\epsilon p\}$  does not depend on the selection of the orthonormal basis. Since $\{k_p\}$ is square root dominated, we get
\begin{equation}
    \sum\limits_{p=1}^{\infty}\mathbb{P}(A_p) = \sum\limits_{p=1}^{\infty}\mathbb{P}(||\underset{L}{\operatorname{proj}}(Z^p_s)||^2>\epsilon p) \leq \frac{3M}{\epsilon^2} \sum\limits_{p=1}^{\infty}\frac{k_p^2}{p^2}<\infty.
\end{equation}
Since the $\epsilon$ was arbitrary, an application of Borel-Cantelli lemma yields,  
$$
\lim\limits_{p\rightarrow \infty}\frac{1}{p}||\underset{L}{\operatorname{proj}}(Z^p_s)||^2=0 \text{ almost surely }.
$$
\end{proof}

\begin{proof}[{Proof of Proposition \ref{t5}}:]
Consider $(h,\underset{L}{\operatorname{proj}}(h))$ and let $\{u_1,u_2,...,u_{k_p}\}$ be an orthonormal basis of $L$. Using (\ref{core}) and setting 
$\epsilon_i^p=\frac{u_i^TZ}{\sqrt{p}}\frac{\phi}{\frac{s}{\sqrt{p}}\sqrt{n}}$,
we obtain
\begin{align}
    (h,\underset{L}{\operatorname{proj}}(h))
    &=\sum\limits_{i=1}^{k_p}(h,u_i)^2 \nonumber\\
    &=\sum\limits_{i=1}^{k_p}(\Gamma_p(b,u_i)+\frac{u_i^TZ}{\sqrt{p}}\frac{\phi}{\frac{s}{\sqrt{p}}\sqrt{n}})^2 \nonumber\\
    &=\Gamma_p^2\sum\limits_{i=1}^{k_p}(b,u_i)^2+2\Gamma_p\sum\limits_{i=1}^{k_p}(b,u_i)\epsilon_i^p+\sum\limits_{i=1}^{k_p}(\epsilon_i^p)^2  \nonumber \\
    &=\Gamma_p^2(b,\underset{L}{\operatorname{proj}}(b))+2\Gamma_p\sum\limits_{i=1}^{k_p}(b,u_i)\epsilon_i^p+\sum\limits_{i=1}^{k_p}(\epsilon_i^p)^2. \label{e26}
\end{align}

 We can relate the third term in (\ref{e26}) to Lemma {\ref{l22}} as follows,
\begin{align}
    \sum\limits_{i=1}^{k_p}(\epsilon_i^p)^2
    &=\frac{1}{n\frac{s_p^2}{p}}\frac{1}{p}\sum\limits_{i=1}^{k_p}\sum\limits_{l,s} \phi_l\phi_s(u_i,Z_l)(u_i,Z_s) \nonumber\\
    &=\frac{1}{n\frac{s_p^2}{p}} \sum\limits_{l,s} \phi_l\phi_s \frac{1}{p}\sum\limits_{i=1}^{k_p} (u_i,Z_l)(u_i,Z_s) \nonumber \\
    &=\frac{1}{n\frac{s_p^2}{p}}\sum\limits_{l, s} \phi_l\phi_s \frac{1}{p}(Z_s,\underset{L}{\operatorname{proj}}(Z_l))\label{new15}.
\end{align}
Using 
the Cauchy-Schwarz inequality yields
\begin{align}
    \frac{1}{p}(Z_s,\underset{L}{\operatorname{proj}}(Z_l)) 
    &= \frac{1}{p}(\underset{L}{\operatorname{proj}}(Z_s),\underset{L}{\operatorname{proj}}(Z_l))\nonumber \\
    &\leq \sqrt{\frac{1}{p}||\underset{L}{\operatorname{proj}}(Z_s)||^2}\sqrt{\frac{1}{p}||\underset{L}{\operatorname{proj}}(Z_l)||^2} \nonumber\\
    &=\sqrt{\frac{1}{p}(Z_s,\underset{L}{\operatorname{proj}}(Z_s))}\sqrt{\frac{1}{p}(Z_l,\underset{L}{\operatorname{proj}}(Z_l))} \label{new15d}
\end{align}

Moreover, since Lemma {\ref{l22}} applies to each column of $Z$ we get,
\begin{equation}
    \lim\limits_{p\rightarrow\infty}\frac{1}{p}(Z_s,\underset{L}{\operatorname{proj}}(Z_l))=0 \label{a9}.
\end{equation}
for any $l$ and $s$. By Lemma \ref{l3}, $s_p^2/p$ is bounded below, so $\frac{1}{n\frac{s_p^2}{p}}$ has a finite limit supremum as $p$ goes to infinity. Also, since the vector $\phi$ is orthonormal and is of finite dimension, we obtain 
\begin{equation}
    \lim\limits_{p\rightarrow \infty} \sum\limits_{i=1}^{k_p}(\epsilon_i^p)^2=0 \label{new20}
\end{equation}
almost surely.
\\\\
Now we are ready to prove the first part of the proposition \ref{t5},
\begin{align}
    &|(h,\underset{L}{\operatorname{proj}}(h))-(h,b)^2(b,\underset{L}{\operatorname{proj}}(b))| \nonumber\\
    &\leq |\Gamma^2_p-(h,b)^2|(b,\underset{L}{\operatorname{proj}}(b))+
     2|\Gamma_p\sum\limits_{i=1}^{k_p}(b,u_i)\epsilon_i^p|+\sum\limits_{i=1}^{k_p}(\epsilon_i^p)^2
    \label{new16}\\
    &=|\Gamma^2_p-(h,b)^2|\sum\limits_{i=1}^{k_p}(b,u_i)^2+2|\Gamma_p\sum\limits_{i=1}^{k_p}(b,u_i)\epsilon_i^p|+\sum\limits_{i=1}^{k_p}(\epsilon_i^p)^2 \label{new17} \\
    &\leq |\Gamma^2_p-(h,b)^2|\sum\limits_{i=1}^{k_p}(b,u_i)^2+2|\Gamma_p|\sqrt{\sum\limits_{i=1}^{k_p}(b,u_i)^2}\sqrt{\sum\limits_{i=1}^{k_p}(\epsilon_i^p)^2}+\sum\limits_{i=1}^{k_p}(\epsilon_i^p)^2 \label{new18}\\
    &\leq |\Gamma^2_p-(h,b)^2|+2|\Gamma_p|\sqrt{\sum\limits_{i=1}^{k_p}(\epsilon_i^p)^2}+\sum\limits_{i=1}^{k_p}(\epsilon_i^p)^2 \label{new19}.
\end{align}
Note that we used the Cauchy-Schwarz and Bessel's inequalities for the transitions from  (\ref{new17}) to (\ref{new18}), and  (\ref{new18}) to (\ref{new19}), respectively. Using (\ref{new20}) and Lemma \ref{l3}, the right hand side of the inequality (\ref{new19}) converges to zero almost surely.

For the second part of the Propostion, a similar argument shows that 
\[
|(b,\underset{L}{\operatorname{proj}}(h))-(h,b)(b,\underset{L}{\operatorname{proj}}(b))|
\]
converges to zero almost surely.

 
 For the last part of Proposition \ref{t5},
 \begin{align*}
     &|| \underset{L}{\operatorname{proj}}(h)-(h,b)\underset{L}{\operatorname{proj}}(b)||^2 \nonumber\\
     &=||\underset{L}{\operatorname{proj}}(h)||^2-2(h,b)(\underset{L}{\operatorname{proj}}(h),\underset{L}{\operatorname{proj}}(b))+||\underset{L}{\operatorname{proj}}(b)||^2 \label{a10} \\
     &=(h,\underset{L}{\operatorname{proj}}(h))-2(h,b)(b,\underset{L}{\operatorname{proj}}(h))+(h,b)^2(b,\underset{L}{\operatorname{proj}}(b)) \\
     &=\big[(h,\underset{L}{\operatorname{proj}}(h))-(h,b)^2(b,\underset{L}{\operatorname{proj}}(b))\big]+2(h,b)\big[(h,b)(b,\underset{L}{\operatorname{proj}}(b))-(b,\underset{L}{\operatorname{proj}}(h))\big] .
 \end{align*}
 An application of the first two parts of Proposition \ref{t5} completes the proof. \end{proof}

\subsection{Proof of Proposition \ref{t7}}

Two preliminary lemmas are useful.

\begin{lem}\label{l1}
Fix an $\epsilon>0$, and positive integers $p, k_p$ with $1 \leq k_p < p$. For $g \in O(n)$, define the linear subspace $H_p(g)=<ge_1,ge_2,...,ge_{k_p}>$ of $\mathbb{R}^p$. Let $u\in \mathbb{S}^{p-1}$ be a fixed vector, and define
\begin{equation}
    N_p^{u}=\{g\in O(p)\big|\text{ } ||\underset{H_p(g)}{\operatorname{proj}}(u)||^2>\epsilon\} \nonumber.
\end{equation}
Similarly, for a fixed subspace $V_p$ of dimension $k_p$, we define
\begin{equation}
    M_p^{V}=\{g\in O(p)\big | \text{ } ||\underset{V_p}{\operatorname{proj}}(ge_1)||^2>\epsilon\} \nonumber.
\end{equation}
For any choice of the nonrandom vector $u$ and the nonrandom subspace $V_p$ we have  $\sigma(N_p^{u})=\sigma(M_p^{V})$, where $\sigma$ denotes Haar measure on $O(p)$.
\end{lem}

The lemma asserts that, under the Haar measure, the event that involves the projection of a fixed point on $\mathbb{S}^{p-1}$ onto a random linear subspace is equivalent to the event that involves the projection of a random point on $\mathbb{S}^{p-1}$ to a fixed linear subspace. 

The proof of this lemma makes use of the fact that Haar measure is invariant under left and right translations, and is omitted.

\begin{lem}\label{l2}
For $\epsilon >0$ and any linear subspace $V_p \subset \mathbb{R}^p$ of dimension $k_p$, define, as before,
\begin{equation}
    M_p^{V}=\{g\in O(p)\big | \text{ } ||\underset{V_p}{\operatorname{proj}}(ge_1)||^2>\epsilon\} \nonumber .
\end{equation}
Then we have the following bound depending only on $\epsilon$, $p$, and $k_p$:
\begin{equation}
    \sigma(M_p^{V}) \leq D_p= k_p \frac{I_{1-\frac{\epsilon}{k_p}}(\frac{p-1}{2},\frac{1}{2})}{I(\frac{p-1}{2},\frac{1}{2})} \nonumber
\end{equation} 
where $I(x,y)$ and $I_{a}(x,y)$ are the beta function and the incomplete beta function, respectively. Moreover, if $k_p$ is of order $O(p^{\alpha})$ for $\alpha<1$, the bound satisfies

\begin{equation}
    \sum\limits_{p=1}^{\infty} D_p < \infty \nonumber.
\end{equation}
\end{lem}

We omit the purely geometric proof, which uses an analysis of volumes of spherical caps and properties of the beta functions.

\begin{proof}[\text{Proof of the Theorem \ref{t7}}.]
For the random linear subspace 
\[
\mathcal{H}_p(\omega)=<\xi(\omega)e_i\text{ } | \text{ } i=1,2...,k_p > ,
\]
where $\xi$ is an $O(n)$-valued random variable inducing the Haar measure on $O(n)$,
 we would like to show
\begin{equation}
    \lim\limits_{p\rightarrow \infty}||\underset{\mathcal{H}_p}{\operatorname{proj}}(z)||^2=0 \text{ }\text{ almost surely.} \nonumber
\end{equation}
It suffices to show, for any $\epsilon>0$ that
\begin{equation}
    \limsup\limits_{p\rightarrow \infty}||\underset{\mathcal{H}_p}{\operatorname{proj}}(z)||^2 \leq \epsilon \text{ }\text{ almost surely.} \label{goal}
\end{equation}

Fix an $\epsilon>0$, and define the set 
\[
F_p=\big\{(g,y)\in O(p)\times \mathbb{S}^{p-1}\big|\text{ }\sum\limits_{i=1}^{k_p} (ge_i,y)^2> \epsilon \big\}.
\]
 Now consider the event $A_p=\{\omega \in \Omega \big | \text{ } (\xi(\omega),z_p(\omega))\in F_p\}$. 
 Since
\begin{equation}
    A_p
    =\{\omega\in \Omega\big| \text{ }  ||\underset{\mathcal{H}_p(\omega)}{\operatorname{proj}}(z_p(\omega)||^2 > \epsilon \},
    \nonumber
\end{equation}
if we show
\[
\sum\limits_{p=1}^{\infty}\mathbb{P}(A_p)<\infty
\]
then an application of the Borel-Cantelli lemma will establish {(\ref{goal})}.

To that end,  note
\begin{align}
    \mathbb{P}(A_p)
    &=\mathbb{E}_{\mathbb{P}}[\mathbf{1}_{\{A_p\}}]=\mathbb{E}[\mathbf{1}_{\{F_p\}}(\xi,z_p)] \nonumber\\
    &=\mathbb{E}[\mathbb{E}[\mathbf{1}_{\{F_p\}}(\xi,z_p)\big|z_p]]. \label{1}
\end{align}
For any $u\in\mathbb{S}^{p-1}$ define the (nonrandom) function $h(u)=\mathbb{E}[\mathbf{1}_{\{F_p\}}(\xi,u)]$. Since $\xi$ is independent of $z_p$, we have for any $\omega\in \Omega$
\begin{equation}
    \mathbb{E}[\mathbf{1}_{\{F_p\}}(\xi,z_p)\big|z_p](\omega)=h(z_p(\omega)), \nonumber
\end{equation}

\noindent
and therefore
$$\mathbb{E}[\mathbb{E}[\mathbf{1}_{\{F_p\}}(\xi,z_p)\big|z_p]]=\mathbb{E}[h(z_p)].$$ 

On the other hand we have $h(u)=\mathbb{P}(\omega\in \Omega \big| \text{ } \xi(\omega)\in N_{p}^u)=\sigma(N_{p}^u)$, using the notation of Lemmas \ref{l1} and \ref{l2}.
By the use of those lemmas we obtain a bound $D_p$ on $h(u)$ that does not depend on $u$. 
Hence, using (\ref{1}), we have
\begin{align*}
    \mathbb{P}(A_p)=\mathbb{E}[h(z_p)]\leq \mathbb{E}[D_p]=D_p, \nonumber
\end{align*}
and by an application of Lemma \ref{l2} we get
\begin{equation}
\sum\limits_{p=1}^{\infty} \mathbb{P}(A_p)\leq \sum\limits_{p=1}^{\infty} D_p <\infty. \nonumber
\end{equation}

\end{proof}



\subsection{Proof of Proposition \ref{p1}}

From now on we will use the shorthand  $\nu$ for the quantity $(b_1,b_2)_{\infty}$. 

Also, recall assumption \textbf{(5)}: $\mu_{\infty}(\beta_1)=\mu_{\infty}(\beta_2)$ and $d_{\infty}(\beta_1) = d_{\infty}(\beta_2)$ a.s. Therefore
\begin{equation}
    \lim\limits_{p\rightarrow\infty}\frac{|\beta|}{\sqrt{p}}=\mu_{\infty}(\beta_1)\sqrt{1+d_{\infty}(\beta_1)^2}=\mu_{\infty}(\beta_2)\sqrt{1+d_{\infty}(\beta_2)^2}=\lim\limits_{p\rightarrow\infty}\frac{|\beta_2|}{\sqrt{p}} \label{i1}
\end{equation}
and
\begin{equation}
    (b_1,q)_{\infty}=\frac{\lim\limits_{p\rightarrow\infty}\sum\limits_{i=1}^p\beta_1(i)}{\frac{|\beta_1|}{\sqrt{p}}}=\frac{\mu_{\infty}(\beta_1)}{\frac{|\beta_1|}{\sqrt{p}}}=\frac{\mu_{\infty}(\beta_2)}{\frac{|\beta_2|}{\sqrt{p}}}=\frac{\lim\limits_{p\rightarrow\infty}\sum\limits_{i=1}^p\beta_2(i)}{\frac{|\beta_2|}{\sqrt{p}}}=(b_2,q)_{\infty}. \label{i2}
\end{equation}


For the proof of Proposition {\ref{p1}} we will need three intermediate lemmas.  
\begin{lem}\label{AppendixC,lem1}
For $x_1,x_2\in\mathbb{R}^n$ set $x=\bigg[\begin{array}{c}
     x_1 \\ 
     x_2
\end{array}\bigg]\in \mathbb{R}^{2n}$ and impose $||x||^2=1$. Recall $\nu = (b_1,b_2)_{\infty}$ and define the following functions:
\begin{equation}
    g_p(x)\equiv g_p(x_1,x_2) =\frac{||Rx||^2}{2np}=\frac{||R_1x_1+R_2x_2||^2}{2np},
\end{equation}
\begin{equation}
 g_{\infty}(x)=\left(\lim\limits_{p\rightarrow\infty}\frac{|\beta|^2}{p}\right)\frac{1}{2n}\bigg[\sum\limits_{t=1}^2(X_t^Tx_t)^2+2\nu(X_1^Tx_1)(X_2^Tx_2)+\delta^2\bigg].
\end{equation}

Then $g_p(x)$ converges to $g_{\infty}(x)$ uniformly almost surely as $p$ tends to $\infty$.
\end{lem}
\begin{proof}
Notice
\begin{equation}
    g_p(x_1,x_2)=\frac{1}{2n}\bigg[\frac{1}{p}\sum\limits_{t=1}^2||R_tx_t||^2+\frac{1}{p}2(R_1x_1,R_2x_2)\bigg]
\end{equation}
By using the proof of Lemma \ref{lemmaA2} on  each summand, the first term in the bracket converges to 
\begin{equation}
\left(\lim\limits_{p\rightarrow\infty}\frac{|\beta|^2}{p}\right)\sum\limits_{t=1}^2(X_t^Tx_t)^2+\delta^2
\end{equation}
uniformly almost surely. Hence it suffices to prove that the remaining term $\frac{1}{np}(R_1x_1,R_2x_2)$ converges to $\left(\lim\limits_{p\rightarrow\infty}\frac{|\beta|^2}{p}\right)\nu\frac{1}{n}(X_1^Tx_1)(X_2^Tx_2)$ uniformly almost surely. We can re-write it as follows,
\begin{align}
    \frac{1}{np}(R_1x_1,R_2x_2)=
    &\frac{1}{n}(X_1^Tx_1)\frac{1}{p}\beta_1^TZ_2x_2+\frac{1}{n}(X_2^Tx_2)\frac{1}{p}\beta_2^TZ_1x_1 \nonumber \\
    +&\frac{1}{n}(X_1^Tx_1)(X_2^Tx_2)\frac{1}{p}\beta_1^T\beta_2+\frac{x_1^TZ_1^TZ_2x_2}{np}.
\end{align}
The first and second term converges to zero uniformly almost surely by an application of Lemma \ref{lemmaA1}  and by the fact that the terms $X_1^Tx_1$ and $X_2^Tx_2$ can be uniformly bounded by $|X_1|$ and $|X_2|$, respectively. Also note the third term converges to the desired limit uniformly almost surely. Hence it is left to prove that the  fourth term converges to 0 uniformly almost surely. We can re-write it as,
\begin{equation}
    \frac{x_1^TZ_1^TZ_2x_2}{np}=\frac{1}{n}\sum\limits_{i,j}^n(x_1(i))(x_2(j))\frac{1}{p}\sum\limits_{k=1}^p(Z_1)_{ki}(Z_2)_{kj}. \label{mixed}
\end{equation}
Lets now fix $i,j$ and set $Y_k:=(Z_1)_{ki}(Z_2)_{kj}$. Since the entries $(Z_1)_{ki}$ and $(Z_2)_{kj}$ belong to the same row of $Z$, they are uncorrelated by the assumption \textbf{4}. Moreover, $\{Y_k\}_{k=1}^{\infty}$ is a sequence of independent random variables by the assumption \textbf{4} as well. Using that and an application of Cauchy-Shawardz inequality along with the 4th moment condition on the entries of $Z$ we get,
\begin{equation}
    \mathbb{E}[Y_k]=\mathbb{E}[(Z_{1})_{ki}]\mathbb{E}[(Z_2)_{kj}]=0, \nonumber
\end{equation}
\begin{equation}
    \mathbb{E}[Y_k^2]=\mathbb{E}[(Z_{1})_{ki}^2(Z_{2})_{kj}^2]
\leq \sqrt{\mathbb{E}[(Z_{1})_{ki}^4]\mathbb{E}[(Z_{2})_{kj}^4]}\leq M. \nonumber
\end{equation}
From here we can apply the Kolmogorov strong law of large numbers to conclude 
\begin{equation}
    \lim\limits_{p\rightarrow\infty}\frac{1}{p}\sum\limits_{k=1}^p(Z_1)_{ki}(Z_2)_{kj}=\lim\limits_{p\rightarrow\infty}\frac{1}{p}\sum\limits_{k=1}^pY_k=0 \text{ } \text{ almost surely} \label{conv}
\end{equation}
which is true for all $i,j\in\{1,2,...,n\}$. We are now ready to complete the argument.
\begin{align*}
\lim\limits_{p\rightarrow\infty}\Big|\frac{1}{n}&\sum\limits_{i,j}^n(x_1)_i(x_2)_j\frac{1}{p}\sum\limits_{k=1}^p(Z_1)_{ki}(Z_2)_{kj}\Big| \nonumber \\
&\leq \Big|\frac{1}{n}\sum\limits_{i,j}^n(x_1)_i(x_2)_j\Big|\lim\limits_{p\rightarrow\infty}\max\limits_{i,j}\Big|\frac{1}{p}\sum\limits_{k=1}^p(Z_1)_{ki}(Z_2)_{kj}\Big|
\nonumber\\
&=\frac{1}{n}\Big|\sum\limits_{i}^n(x_1)_i\Big|\Big|\sum\limits_{i}^n(x_2)_i\Big|\lim\limits_{p\rightarrow\infty}\max\limits_{i,j}\Big|\frac{1}{p}\sum\limits_{k=1}^p(Z_1)_{ki}(Z_2)_{kj}\Big| \nonumber\\
&\leq \sqrt{\sum\limits_{i}^n(x_1)_i^2}\sqrt{\sum\limits_{i}^n(x_2)_i^2}\lim\limits_{p\rightarrow\infty}\max\limits_{i,j}\Big|\frac{1}{p}\sum\limits_{k=1}^p(Z_1)_{ki}(Z_2)_{kj}\Big|\nonumber \\
&=\lim\limits_{p\rightarrow\infty}\max\limits_{i,j}\Big|\frac{1}{p}\sum\limits_{k=1}^p(Z_1)_{ki}(Z_2)_{kj}\Big|=0 \text{ }\text{ almost surely}.
\end{align*}
The final limit is $0$ since number of possible pairs of $i$ and $j$ is $n^2<\infty$ and we have {(\ref{conv})}. 
\end{proof}

The proof of the following Lemmas is a straightforward computation.

\begin{lem}\label{AppendixC,lem2}
Let
\begin{equation}
     \lambda=\frac{1}{2}\bigg((|X_1|^2+|X_2|^2)+\sqrt{(|X_1|^2-|X_2|^2)^2+4\nu^2|X_1|^2|X_2|^2}\bigg). \nonumber
\end{equation}

The maximum of the function $g_{\infty}(x)\equiv g_{\infty}(x_1,x_2)$ defined in Lemma \ref{AppendixC,lem1} is attained at 
\begin{equation}
    x^*=\bigg[\begin{array}{c}
         x_1^*  \\
         x_2^* 
    \end{array}\bigg]
    = \pm
    \bigg[\begin{array}{c}
         \frac{X_1}{|X_1|}a_1  \\
         \frac{X_2}{|X_2|}a_2 
    \end{array}\bigg]
\end{equation}
where
\\
\[\centering
\begin{tabular}{cc}
 $a_1=\sqrt{\frac{\lambda-|X_2|^2}{2\lambda-|X|^2}}$,   $a_2=\sqrt{\frac{\lambda-|X_1|^2}{2\lambda-|X|^2}}$ & for $\nu > 0$  \\
$a_1=\frac{1}{\sqrt{2}}$, $a_2=\frac{1}{\sqrt{2}}$ & for $\nu=0$ and $|X_2|=|X_1|$ \\
$a_1=1$, $a_2$=0 & for $\nu=0$ and  $|X_1|>|X_2|$ \\
$a_1=0$, $a_2=1$ & for $\nu=0$ and $|X_2|>|X_1|$ 
\end{tabular}
\]
\\
and the maximum value is
\begin{equation}
     max(g_{\infty}(x_1,x_2))=  \frac{1}{2n}\Big[\Big(\lim\limits_{p\rightarrow\infty}\frac{|\beta|^2}{p}\Big)\lambda+\delta^2\Big]
\end{equation}
\end{lem}

\begin{lem}\label{AppendixC,lem3}
With the notation as before and $a_1$, $a_2$ and $\lambda$ as defined as in Lemma \ref{AppendixC,lem2},
\begin{equation}
   \lim\limits_{p\rightarrow\infty}\frac{s^2}{p}=\frac{1}{2n}(\Big(\lim\limits_{p\rightarrow\infty}\frac{|\beta|^2}{p}\Big)\lambda+\delta^2)  \text{ }\text{ and }\text{ }  
   \lim\limits_{p\rightarrow\infty}\chi=\bigg[\begin{array}{c}
         \frac{X_1}{|X_1|}a_1  \\
         \frac{X_2}{|X_2|}a_2 
    \end{array}\bigg]\nonumber
\end{equation}
almost surely.
\end{lem}
\begin{proof}
Define  $\chi_1,\chi_2\in\mathbb{R}^n$  by
$\chi=\bigg[\begin{array}{c}
     \chi_1 \\
     \chi_2
\end{array}\bigg]$. The definitions of $s$,$h$ and $\chi$ are such that
\begin{equation}
    sh=\frac{R\chi}{\sqrt{2n}} \label{AppendixC,0}
\end{equation}
Since $s^2$ is the largest eigenvalue we get 
\begin{equation}
    \frac{s^2}{p}=\frac{|Rx|^2}{2np}
    =g_p(\chi_1,\chi_2)=\underset{|x_1|^2+|x_2|^2=1}{sup}g_{p}(x_1,x_2). \label{a29}
\end{equation}
By Lemma {\ref{AppendixC,lem1}}, we have the uniform almost sure convergence of $g_p(x_1,x_2)$ to $g_{\infty}(x_1,x_2)$. Hence the supremum and limits are interchangeable:
\begin{align}
    \lim\limits_{p\rightarrow\infty} g_p(\chi_1,\chi_2)&=\lim\limits_{p\rightarrow\infty}\underset{|x_1|^2+|x_2|^2=1}{\sup}g_{p}(x_1,x_2) \nonumber \\
    &=\underset{|x_1|^2+|x_2|^2=1}{\sup}\lim\limits_{p\rightarrow\infty}g_{p}(x_1,x_2)=\underset{|x_1|^2+|x_2|^2=1}{\sup}g_{\infty}(x_1,x_2)  \nonumber \\
    &=g_{\infty}(x_1^*,x_2^*). \label{a27}
\end{align}
\\   
By ({\ref{a27}}) along with the Lemma {\ref{AppendixC,lem2}},
\begin{equation}
    \lim\limits_{p\rightarrow\infty}\frac{s^2}{p}= \lim\limits_{p\rightarrow\infty}g_p(\chi_1,\chi_2)=g_{\infty}(x_1^*,x_2^*)=\frac{1}{2n}(\Big(\lim\limits_{p\rightarrow\infty}\frac{|\beta|^2}{p}\Big)\lambda+\delta^2).
\end{equation}
This proves the first part of the Lemma. On the other hand, the following almost sure convergence also follows from Lemma {\ref{AppendixC,lem1}}:
\begin{equation}
    \lim\limits_{p\rightarrow\infty} |g_p(\chi_1,\chi_2)-g_{\infty}(\chi_1,\chi_2)|=0. \label{a28}
\end{equation}
Combining ({\ref{a27}}) and ({\ref{a28}}) we obtain, 
\begin{equation}
    \lim\limits_{p\rightarrow\infty} |g_{\infty}(\chi_1,\chi_2)-g_{\infty}(x_1^*,x_2^*)|=0 \label{AppendixC,16}
\end{equation}
Expanding equation \ref{AppendixC,0} yields
\begin{equation}
    (h,q)=\frac{(b_1,q)|\beta_1|X_1^T\chi_1+(b_2,q)|\beta_1|X_1^T\chi_1+q^TZ_1\chi_1+q^TZ_2\chi_2}{s\sqrt{2n}} \label{AppendixC,14}. 
\end{equation}
By an application of Lemma \ref{lemmaA1}, it is straightforward to show that the last two terms vanish almost surely.
Hence we have,
\begin{equation}
    (h,q)_{\infty}=(b_1,q)_{\infty}\bigg(\lim\limits_{p\rightarrow \infty}\frac{|\beta_1|X_1^T\chi_1}{s\sqrt{2n}}\bigg)+(b_2,q)_{\infty}\bigg(\lim\limits_{p\rightarrow \infty}\frac{|\beta_2|X_2^T\chi_2}{s\sqrt{2n}}\bigg). \label{AppendixC,17}
\end{equation}
Note that the sequence of points $\chi=\bigg[\begin{array}{c}
     \chi_1 \\
     \chi_2
\end{array}\bigg]$ lie on a closed and bounded set $\mathbb{S}^{2n-1}$. Hence for any sub-sequence,  a further sub-sequence converges. Since we have (\ref{AppendixC,16}) and $g_{\infty}$ is continuous, those sub-sequence of vectors must converge to either $\bigg[\begin{array}{c}
         \frac{X_1}{||X_1||}a_1  \\
         \frac{X_2}{|X_2|}a_2 
    \end{array}\bigg]$ or $-\bigg[\begin{array}{c}
         \frac{X_1}{||X_1||}a_1  \\
         \frac{X_2}{|X_2|}a_2 
    \end{array}\bigg]$. 

By (\ref{AppendixC,17}) the fact that $a_1\geq0$, $a_2\geq 0$, and $(h,q) \geq 0$, there must be a further sub-sequence that converges to $$\bigg[\begin{array}{c}
         \frac{X_1}{||X_1||}a_1  \\
         \frac{X_2}{|X_2|}a_2 
    \end{array}\bigg].$$ 
Hence
\begin{equation}
    \lim\limits_{p\rightarrow\infty} \chi_1=\frac{X_1}{||X_1||}a_1 \text{ }\text{ and }\text{ } \lim\limits_{p\rightarrow\infty} \chi_2=\frac{X_2}{||X_2||}a_2 \label{a30}.
\end{equation}
\end{proof}
Now we are ready at last to prove Proposition {\ref{p1}}.
\begin{proof}[\text{Proof of Proposition \ref{p1}}:]
Let $(h_1,s_1)$ and $(h_2,s_2)$ be the leading eigenpairs for the sample covariance matrices $S_1=\frac{1}{n}R_1R_1^T$ and $S_2=\frac{1}{n}R_2R_2^T$ respectively. Also let $\phi_1$ and $\phi_2$ be the right singular vectors of $R_1$ and $R_2$ that are associated with $h_1$ and $h_2$ respectively. Hence we can write 
\begin{equation}
    h_1=\frac{R_1\phi_1}{s_1\sqrt{n}}=\frac{\beta_1X_1^T\phi_1+Z_1\phi_1}{s_1\sqrt{n}}
\end{equation}
\begin{equation}
    h_2=\frac{R_2\phi_2}{s_2\sqrt{n}}=\frac{\beta_2X_2^T\phi_2+Z_2\phi_2}{s_2\sqrt{n}}
\end{equation}
\begin{equation}
    h=\frac{R\chi}{s\sqrt{2n}}=\frac{\beta_1X_1^T\chi_1+Z_1\chi_1+\beta_2^TX_2\chi_2+Z_2\chi_2}{s\sqrt{2n}}
\end{equation}
Now define
\begin{equation}
    \tilde{h}=\frac{s_1}{s\sqrt{2}}a_1h_1+\frac{s_2}{s\sqrt{2}}a_2h_2 \label{eq:tildeh}.
\end{equation}
Clearly, $\tilde{h}$ resides in the span of $h_1$ and $h_2$.
By  Lemma \ref{lemmaA2} we have $\lim\limits_{p\rightarrow\infty} \phi_1=\frac{X_1}{|X_1|}$ and $\lim\limits_{p\rightarrow\infty} \phi_2=\frac{X_2}{|X_2|}$ almost surely. We have,
\begin{align}
    ||\tilde{h}-h||
    &=\bigg|\bigg|\frac{\beta_1X_1^T}{s\sqrt{2n}}(a_1\phi_1-\chi_1)+\frac{\beta_2X_2^T}{s\sqrt{2n}}(a_2\phi_2-\chi_2) \nonumber\\
    &+\frac{Z_1}{s\sqrt{2n}}(a_1\phi_1-\chi_1)+\frac{Z_2}{s\sqrt{2n}}(a_2\phi_2-\chi_2)\bigg|\bigg| \nonumber\\
    &\leq \bigg\{\bigg|\bigg|\frac{\beta_1X_1^T}{s\sqrt{2n}}\bigg|\bigg|_F+\bigg|\bigg|\frac{Z_1}{s\sqrt{2n}}\bigg|\bigg|_F\bigg\}\bigg|\bigg|a_1\phi_1-\chi_1\bigg|\bigg| \nonumber\\
    &+\bigg\{\bigg|\bigg|\frac{\beta_2X_2^T}{s\sqrt{2n}}\bigg|\bigg|_F+\bigg|\bigg|\frac{Z_2}{s\sqrt{2n}}\bigg|\bigg|_F\bigg\}\bigg|\bigg|a_2\phi_2-\chi_2\bigg|\bigg|.
\end{align}
By Lemmas \ref{AppendixC,lem1}  and  \ref{lemmaA2},  both terms $a_1\phi_1-\chi_1$ and $a_2\phi_2-\chi_2$ converge to $0$ almost surely. Since we have $\lim\limits_{p\rightarrow\infty}\frac{s}{\sqrt{p}}\in (0,\infty)$ from the proof of \ref{AppendixC,lem1}, a few applications of strong law of large numbers shows that the Frobenious norm terms remain finite in the asymptotic regime. This completes the proof.  
\end{proof}

\subsection{Proof of Theorem \ref{T4}(b)}

 We first need to tackle the following technical propositions.
\begin{prop} \label{p2}
Under assumptions \textbf{1-5},
\begin{enumerate}
    \item $|\nu-(b,q)_{\infty}^2|>0$ almost surely if and only if $|\rho_{\infty}(\beta_1,\beta_2)|>0$ almost surely,
    \item $\nu<1$ almost surely if and only if $\rho_{\infty}(\beta_1,\beta_2)<1$ almost surely, and
    \item $1+\nu-2(b,q)^2_{\infty}>0$ almost surely if and only if  $\rho_{\infty}(\beta_1,\beta_2)>-1$ almost surely.
\end{enumerate}
\end{prop}
\begin{proof}
From the definitions of $d_p(\beta_1)$, $d_p(\beta_1)$ and $d_p(\beta_1,\beta_2)$  it follows that
\begin{equation}
    (b_1,b_2)-(b_1,q)(b_2,q)=
\frac{d_p(\beta_1,\beta_2)}{\sqrt{1+d_p(\beta_1)^2}\sqrt{1+d_p(\beta_2}^2)} \label{AppendixC,ee4}
\end{equation}
\begin{equation}
    (b_1,q)=\frac{1}{\sqrt{1+d_p(\beta_1)^2}} \text{ }\text{ and }\text{ } (b_2,q)=\frac{1}{\sqrt{1+d_p(\beta_2)^2}}. \label{AppendixC,ee5}
\end{equation}
Now using ({\ref{AppendixC,ee4}}) and the assumption \textbf{(5)} we can prove the first part of the proposition as,
\begin{align}
    &|\nu-(b,q)_{\infty}^2|
    =\lim\limits_{p\rightarrow} |(b_1,b_2)-(b_1,q)(b_2,q)| \nonumber\\
    &=|\frac{d_{\infty}(\beta_1,\beta_2)}{1+d_{\infty}(\beta)^2}|=|\rho_p(\beta_1,\beta_2)|\frac{d_{\infty}(\beta)^2}{1+d_{\infty}(\beta)^2}
\end{align}

Next, using equations ({\ref{AppendixC,ee4}}), ({\ref{AppendixC,ee5}}) and the assumption \textbf{(5)} we can re-write $\nu$ as,
\begin{align}
    \nu=\frac{d_{\infty}(\beta_1,\beta_2)}{1+d_{\infty}(\beta)^2}+\frac{1}{1+d_{\infty}(\beta)^2}=\frac{1+d_{\infty}(\beta_1,\beta_2)}{1+d_{\infty}(\beta)^2}
\end{align}
From there it is easy to see
\begin{equation}
    \nu<1 \iff d_{\infty}(\beta_1,\beta_2)<d_\infty(\beta)^2 \iff \rho_{\infty}(\beta)^2<1
\end{equation}
which proves the second assertion. Finally, using the equations ({\ref{AppendixC,ee4}}), ({\ref{AppendixC,ee5}}) and the assumption \textbf{(5)} we can rewrite $1+\nu-2(b,q)^2_{\infty}$ as,
\begin{equation}
    1+\nu-2(b,q)^2_{\infty}=1+\frac{d_{\infty}(\beta_1,\beta_2)}{1+d_{\infty}(\beta)^2}-\frac{1}{1+d_{\infty}(\beta)^2}=\frac{d_{\infty}(\beta)^2+d_\infty(\beta_1,\beta_2))}{1+d_{\infty}(\beta)^2}.
\end{equation}
Therefore
\begin{equation}
    1+\nu-2(b,q)^2_{\infty}>0 \iff d_{\infty}^2(\beta)>-d_{\infty}(\beta_1,\beta_2) \iff \rho_{\infty}(\beta_1,\beta_2)>-1,
\end{equation}
which proves the third part of the proposition. 
\end{proof}

\begin{prop} \label{p3}
Given the modeling assumptions and $|\rho_{\infty}(\beta_1,\beta_2)|>0$ we have 
\begin{equation}
\lim\limits_{p\rightarrow \infty}\big( ||\underset{<q,h_1,h_2>}{\operatorname{proj}}(b_2)||^2-||\underset{<q,h_2>}{\operatorname{proj}}(b_2)||^2
    \big)>0 \text{ } \text{ almost surely. } \nonumber
\end{equation}
\end{prop}

\begin{proof}
First re-write it as
\begin{align}
     ||\underset{<q,h_1,h_2>}{\operatorname{proj}}(b_2)||^2-||\underset{<q,h_2>}{\operatorname{proj}}(b_2)||^2 
     &=||\underset{<q,h_2,h_1^*>}{\operatorname{proj}}(b_2)||^2-||\underset{<q,h_2>}{\operatorname{proj}}(b_2)||^2=(h_1^*,b)^2 \label{a34} 
\end{align}
where $h_1^*=\frac{h_1-\underset{<q,h_2>}{\operatorname{proj}}(h_1)}{||h_1-\underset{<q,h_2>}{\operatorname{proj}}(h_1)||}$. From ({\ref{a34}}) it is sufficient to prove,
\begin{equation}
    (h_1^*,b)_{\infty}^2=\lim\limits_{p\rightarrow\infty}\Big(h_1-\underset{<q,h_2>}{\operatorname{proj}}(h_1),b_2\Big)^2\frac{1}{||h_1-\underset{<q,h_2>}{\operatorname{proj}}(h_1)||^2}>0.
\end{equation}
With some computation and Proposition {\ref{t5}} one can verify that
\begin{equation}
    \lim\limits_{p\rightarrow\infty} \Big|\Big( h_1-\underset{<q,h_2>}{\operatorname{proj}}(h_1),b_2\Big)\Big|=\Big|\frac{(\nu-(b,q)_{\infty}^2)(h_1,b_1)_{\infty}(1-(h_2,b_2)_{\infty}^2)}{1-(h_2,q)^2_{\infty}}\Big|.
\end{equation}
By Lemma \ref{lemmaA2}, we have $(h_1,b_1)_{\infty},(h_2,b_2)_{\infty}\in (0,1)$. By part (1) of  Proposition \textbf{\ref{p2}} we have $|\nu-(b,q)_{\infty}^2|>0$ almost surely. These all together prove
\begin{equation}
   \lim\limits_{p\rightarrow\infty} \Big|\Big(h_1-\underset{<q,h_2>}{\operatorname{proj}}(h_1),b_2\Big)\Big|>0,
\end{equation}
which implies $(h_1^*,b)_{\infty}^2>0$. This finishes the proof. 

\end{proof}

\begin{prop} \label{p4}
Given the modeling assumptions and $0<|\rho_{\infty}(\beta_1,\beta_2)|<1$ almost surely,  we have 
\begin{equation}
\lim\limits_{p\rightarrow \infty}\big( ||\underset{<q,h_1,h_2>}{\operatorname{proj}}(b_2)||^2-||\underset{<q,\Tilde{h}>}{\operatorname{proj}}(b_2)||^2
    \big)>0 \text{ } \text{ almost surely. } \label{E6}
\end{equation}
where $\Tilde{h}$ is defined by equation (\ref{eq:tildeh}).
\end{prop}

\begin{proof}
Rewrite the definition of $\tilde{h}$ as
\begin{equation}
    \Tilde{h}=\frac{s_1}{s\sqrt{2}}a_1h_1+\frac{s_2}{s\sqrt{2}}a_2h_2=A_1h_1+A_2h_2. 
    \label{E3}
\end{equation}
where we set $A_1=\frac{s_1}{s\sqrt{2}}a_1$ and $A_2=\frac{s_2}{s\sqrt{2}}a_2$. We will use the notation $A_{1,\infty}$ and $A_{2,\infty}$ for the limits of $A_1$ and $A_2$ respectively. From the definition of $a_1$ and $a_2$ we have $A_1$ and $A_2$ are nonzero as long as $\nu\neq0$. From the statement of Lemma \ref{AppendixC,lem2} and a use of lemma \ref{AppendixC,lem3} it is easy to recover the following implications of the sub-cases of $\nu=0$, 
\begin{enumerate}[(a)]
    \item If $|X_2|=|X_1|$ and $\nu=0$ then $A_1=A_2=\frac{1}{\sqrt{2}}$,
    \item if $|X_2|>|X_1$ and $\nu=0$ then $A_1=0$, $A_2=1$, 
    \item if $|X_1|>|X_2|$ and $\nu=0$ then $A_1=1$, $A_2=0$.
\end{enumerate}
For the sub-case {(b)} we get $\tilde{h}=h_2$. Hence the assertion of the proposition for this sub-case is same with the assertion of Proposition {\ref{p3}}. Therefore the untreated cases are the sub-cases {(a)}, {(c)} and the case $\nu\neq 0$. For all of them $A_1>0$ and hence  $(h_1,h_2) = (\tilde{h},h_2)$. For that reason we can re-write ({\ref{E6}}) as follows, 
\begin{align}
    &||\underset{<q,h_1,h_2>}{\operatorname{proj}}(b_2)||^2-||\underset{<q,\Tilde{h}>}{\operatorname{proj}}(b_2)||^2 
    =||\underset{<q,\Tilde{h},h_2>}{\operatorname{proj}}(b_2)||^2-||\underset{<q,\Tilde{h}>}{\operatorname{proj}}(b_2)||^2=
   (h_2^*,b_2)^2 \label{E29}
\end{align}
where we set $h_2^*=\frac{h_2-\underset{<q,\tilde{h}>}{\operatorname{proj}(h_2)}}{||h_2-\underset{<q,\tilde{h}>}{\operatorname{proj}(h_2}||}$, and which is by definition orthogonal to the linear subspace generated by $q$ and $\tilde{h}$. Continuing from (\ref{E29}), it is sufficient to show
\begin{equation}
    (h_2^*,b_2)_{\infty}^2=\lim\limits_{p\rightarrow\infty}\Big(\frac{h_2-\underset{<q,\tilde{h}>}{\operatorname{proj}(h_2)}}{||h_2-\underset{<q,\tilde{h}>}{\operatorname{proj}(h_2}||},b_2\Big)^2>0 \text{ }\text{ almost surely}. \nonumber
\end{equation}

By means of Theorem \ref{thm1} and Proposition \ref{p1} we can derive the following decomposition:
\begin{lem} \label{lemma:Qcomputations}
\begin{equation}
            \lim\limits_{p\rightarrow\infty}\big(h_2-\underset{<q,\tilde{h}>}{\operatorname{proj}(h_2)},b_2\big) \nonumber \\
            =\frac{Q_1+(1-(b,q)^2_{\infty})A_{1,\infty}Q_2+Q_3}{1-(\tilde{h},q)_{\infty}^2} \label{AppendixC,ee3}
\end{equation}
where
\begin{align*}
            Q_1&=(b,q)^2_{\infty}(1-\nu)A_{1,\infty}A_{2,\infty}(h_1,b_1)_{\infty}(1-(h_2,b_2)_{\infty}^2) \nonumber \\
            Q_2&=
            A_{1,\infty} (h_2,b_2)(1-(h_1,b_1)^2)- \nu A_{2,\infty}(h_1,b_1)_{\infty}(1-(h_2,b_2)_{\infty}^2) \nonumber \\
            Q_3&=(1-\nu)(1+\nu-2(b,q)^2_{\infty})A_{1,\infty}^2(h_1,b_1)_{\infty}^2(h_2,b_2)_{
            \infty}
\end{align*}
\end{lem}

The proof of Lemma \ref{lemma:Qcomputations} requires some algebraic computations and is omitted here, but complete details appear in \cite{gurdogan2021}.

 Now we will prove  $Q_2\geq 0$, $ Q_1\geq 0$,  and $Q_3>0$ in an orderly fashion. We will use the following implications of the second and the third part of Proposition {\ref{p2}},
\begin{equation}
\nu<1 \text{ }\text {and} \text{ } 1+\nu-2(b,q)^2>0 \text{ }\text{ almost surely.} \label{AppendixC,ee0}
\end{equation}
Lets start with proving $Q_2\geq 0$. Using Lemmas {\ref{AppendixC,lem2}}, {\ref{AppendixC,lem3}} and Lemma \ref{lemmaA2} one can derive the following,
\begin{equation}
    Q_2=\frac{\sqrt{\gamma}\delta^2}{C}\frac{\sqrt{\lambda-|X_2|^2}}{|X_2|}(|X|^2-\lambda) \label{AppendixC,ee1}
\end{equation}
\begin{equation}
    C=\sqrt{(\gamma\lambda+\delta^2)(\gamma|X_1|^2+\delta^2)(\gamma|X_2|^2+\delta^2)(2\lambda-|X|^2)} \text{,}\text{ }\text{ }\gamma=\lim\limits_{p\rightarrow}\frac{|\beta|^2}{p}. \nonumber
\end{equation}
From the definition of $\lambda$ in Lemma {\ref{AppendixC,lem2}}, we can immediately infer that $$max(|X_1|^2,|X_2|^2) \leq\lambda \leq |X|^2.$$ By the assumptions \textbf{(1)}, \textbf{(4)} the remaining terms on the numerator at {\ref{AppendixC,ee1}} are positive. Hence we get $Q_2\geq 0$ almost surely.     
\\\\
In regards to $Q_1$, we have the terms $A_1$, $A_2$ non-negative by their definition. Also, the terms $(h_1,b_1)_{\infty}$, $1-(h_2,b_2)_{\infty}^2$ are positive by Lemma \ref{lemmaA2}. We have $(b,q)_{\infty}^2>0$ positive by a straight forward implication of the modeling assumption \textbf{(1)}. Lastly $1-\nu>0$ by {(\ref{AppendixC,ee0})}. These all together proves that $Q_1\geq 0$.
\\\\
Now, lets see that all terms involve in $Q_3$ are positive. The terms $1-\nu$ and $1+\nu-(b,q)_{\infty}^2$ are positive by {\ref{AppendixC,ee0}}. Also the terms $(h_1,b_1)_{\infty}$ and $(h_2,b_2)_{\infty}$ are positive by Lemma \ref{lemmaA2}. Finally, the term $A_1>0$ for the remaining case/sub-cases being treated. These all-together shows $Q_3>0$.
\\\\ 
Having $Q_1\geq 0$, $Q_2\geq 0$ and $Q_3>0$ and $A_{1,\infty}>0$ on {(\ref{AppendixC,ee3})} proves that,
\begin{equation}
     \lim\limits_{p\rightarrow\infty}\big(h_2-\underset{<q,\tilde{h}>}{\operatorname{proj}(h_2)},b_2\big)>0 
\end{equation}
almost surely. This finishes the proof. 
\end{proof}

\begin{proof}[Proof of Theorem \ref{T4}(b)]

Using Theorem \ref{T1} it suffices to prove
\begin{equation}
    \lim\limits_{p\rightarrow \infty}\big(||h_L-b||-||h^s_q-b||\big)<0
    \text{ and }\text{ } \lim\limits_{p\rightarrow \infty}\big(||h_L-b||-||h_q^d-b||\big)<0 
    \text{ }\text{ almost surely} \label{E1}
\end{equation}
Recalling the definition of $h_L$ and $h_q$ we can re-write it as
\begin{equation}
    \lim\limits_{p\rightarrow \infty}\big( ||\underset{<q,h_1,h_2>}{\operatorname{proj}}(b_2)||^2-||\underset{<q,h>}{\operatorname{proj}}(b_2)||^2
    \big)>0
\end{equation} 
and
\begin{equation}
\lim\limits_{p\rightarrow \infty}\big( ||\underset{<q,h_1,h_2>}{\operatorname{proj}}(b_2)||^2-||\underset{<q,h_2>}{\operatorname{proj}}(b_2)||^2\big)>0 \label{E2}
\end{equation}
almost surely.
Recall from Proposition \ref{p1} that $\Tilde{h}$ converges to $h$ almost surely in the $l_2$ norm. Using this we can update ({\ref{E2}}) with the following equivalent version,
\begin{equation}
    \lim\limits_{p\rightarrow \infty}\big( ||\underset{<q,h_1,h_2>}{\operatorname{proj}}(b_2)||^2-||\underset{<q,\Tilde{h}>}{\operatorname{proj}}(b_2)||^2
    \big)>0
\end{equation}
and
\begin{equation}
\lim\limits_{p\rightarrow \infty}\big( ||\underset{<q,h_1,h_2>}{\operatorname{proj}}(b_2)||^2-||\underset{<q,h_2>}{\operatorname{proj}}(b_2)||^2\big)>0 \label{E4}
\end{equation}
almost surely. Propositions ({\ref{p3}}) and ({\ref{p4}}) finish the proof. 
\end{proof}

\subsection{Two Lemmas}

For the reader's convenience, we state two lemmas from \cite{goldberg2018} that are used above. The reader may consult that paper for proofs.

\begin{lem}[Lemma A.1]\label{lemmaA1}
Let $\eta = \{\eta(i)\}_{i=1}^{\infty}$ be a real sequence with $\mu_p(\eta) = \frac{1}{p} \sum_{i=1}^p \eta(i)$ satisfying
\[
\liminf_{p \to \infty} \mu_p(\eta) > 0.
\]
Let $\{Z(i)\}_{i=1}^{\infty}$ be a sequence of mean zero, pairwise independent and identically distributed real random variables with finite variance.

Then
\[
\frac{\eta_p^T Z_p}{\sqrt{p} |\eta_p|} \to 0
\]
almost surely as $\ \to \infty$, where 
\[\eta_p =(\eta(1), \dots, \eta(p))^T \text{ and } Z_p = (Z(1),\dots,Z(p))^T.
\]
\end{lem}

With the notation and assumptions of our theorems, define a non-degenerate random variable $\sigma_X$ by
\[
\sigma_X^2 = \frac{|X|^2}{n}\mu_{\infty}^2(1+d_{\infty}^2(\beta)),
\]
 and recall: $s_p^2$ is the leading eigenvalue of the sample covariance matrix $S = RR^T/n$, $\ell_p^2$ is the average of the remaining eigenvalues, and $\chi_p$ is the normalized right singular vector of $Y/\sqrt{n}$
corresponding to the singular value $s_p \geq 0$. 
\begin{lem}[Lemma A.2]\label{lemmaA2}
Almost surely as $p \to \infty$,
\[
\frac{s_p}{\sqrt{p}} \to \sqrt{\sigma_X^2 + \delta^2/n}, \, \chi_p \to \frac{X}{|X|}, \text{ and } \frac{\ell_p^2}{p} \to \frac{\delta^2}{n}.
\]
\end{lem}